\documentclass[12pt,letterpaper]{article}
\usepackage{amssymb,amsmath}
\usepackage{graphicx}
\usepackage{enumitem} 
\usepackage{natbib}
\usepackage{accents} 
\newcommand{\blind}{0}

\newcommand{\ave}{\mathop{\mbox{ave}}}

\newcommand{\bone}{\boldsymbol 1}
\newcommand{\bzero}{\boldsymbol 0}
\newcommand{\eps}{\varepsilon}
\newcommand{\gs}{\geqslant}
\newcommand{\ls}{\leqslant}

\newcommand{\bm}{\boldsymbol m}

\newcommand{\bt}{\boldsymbol t}
\newcommand{\bv}{\boldsymbol v}

\newcommand{\bx}{\boldsymbol x}

\newcommand{\bhx}{\boldsymbol{\hat{x}}}

\newcommand{\hlam}{\hat \lambda}
\newcommand{\bmu}{\boldsymbol \mu}

\newcommand{\bL}{\boldsymbol L}

\newcommand{\bP}{\boldsymbol P}

\newcommand{\bR}{\boldsymbol R}
\newcommand{\bS}{\boldsymbol S}

\newcommand{\bT}{\boldsymbol T}

\newcommand{\bU}{\boldsymbol U}

\newcommand{\bhX}{\boldsymbol{\hat X}}

\newcommand{\bX}{\boldsymbol X}

\newcommand{\bZ}{\boldsymbol Z}
\newcommand{\bhZ}{\boldsymbol{\hat Z}}
\newcommand{\bSigma}{\boldsymbol \Sigma}
\newcommand{\bmE}{\boldsymbol{\mathcal{E}}}
\newcommand{\bmP}{\boldsymbol{\mathcal{P\!}}}
\newcommand{\bmT}{\boldsymbol{\mathcal{T\!}}}

\def\mcd{\text{\tiny MCD}}
\def\lmcd{m_{\mcd}}
\def\smcd{s_{\mcd}}
\def\OD{\text{OD}}
\def\SD{\text{SD}}
\def\od{\text{\tiny OD}}
\def\sd{\text{\tiny SD}}

\newcommand{\bcx}{\accentset{\circ}{\bx}}
\newcommand{\bchx}{\boldsymbol{\hat{\accentset{\circ}{x}}}}
\newcommand{\cx}{\accentset{\circ}{x}}
\newcommand{\chx}{\hat{\accentset{\circ}{x}}}
\newcommand{\bcX}{\accentset{\circ}{\bX}}
\newcommand{\bchX}{\boldsymbol{\hat{\accentset{\circ}{X}}}}
\newcommand{\bcZ}{\accentset{\circ}{\bZ}}

\newcommand{\ct}{\accentset{\circ}{t}}
\newcommand{\bcT}{\accentset{\circ}{\bT}}
\newcommand{\cod}{\accentset{\circ}{\text{\footnotesize OD}}}
\newcommand{\csd}{\accentset{\circ}{\text{\footnotesize SD}}}

\newcommand{\sbullet}{
  \hbox{\fontfamily{lmr}\fontsize{4}{0}\selectfont\textbullet}}
\newcommand{\fx}{\accentset{\sbullet}{x}}		
\newcommand{\bfx}{\accentset{\sbullet}{\bx}}
\newcommand{\hfx}{\hat{\accentset{\sbullet}{x}}}		
\newcommand{\bfhx}{\boldsymbol{\hat{\accentset{\sbullet}{x}}}}	
\newcommand{\bfX}{\accentset{\sbullet}{\bX}}
\newcommand{\bfhX}{\boldsymbol{\hat{\accentset{\sbullet}{X}}}}	
\newcommand{\bfT}{\accentset{\sbullet}{\bT}}	
\newcommand{\fod}{\accentset{\sbullet}{\text{\footnotesize OD}}}	

\newcommand{\btx}{\boldsymbol{\tilde x}}

\newcommand{\btX}{\boldsymbol{\tilde X}}

\newcommand{\btT}{\boldsymbol{\widetilde T}}
\newcommand{\tz}{\tilde z}	
		
\newcommand{\btZ}{\boldsymbol{\tilde Z}}

\def\spacingset#1{\renewcommand{\baselinestretch}
{#1}\small\normalsize} \spacingset{1}

\begin{document}
\spacingset{1.45} 
\thispagestyle{empty}

\if0\blind
{ \vspace{-2cm}
	\title{\Large{\bf MacroPCA: An all-in-one PCA
	       method allowing for missing values as 
				 well as cellwise and rowwise outliers}}
  \author{Mia Hubert, Peter J. Rousseeuw, 
          Wannes Van den Bossche\\ 
          \normalsize{Department of Mathematics, 
		      KU Leuven, Belgium}}
  \date{December 9, 2018} 
  \maketitle
} \fi

\if1\blind
{
  \bigskip
  \bigskip
  \bigskip
  \begin{center}
    {\LARGE\bf  MacroPCA: An all-in-one PCA
	   method allowing for missing values as 
		 well as cellwise and rowwise outliers}\\
    \medskip
	  December 9, 2018 
	\end{center}
	\medskip
} \fi

\begin{abstract}
Multivariate data are typically represented by a 
rectangular matrix (table) in which the rows are the 
objects (cases) and the columns are the variables 
(measurements). 
When there are many variables one often reduces the
dimension by principal component analysis (PCA),
which in its basic form is not robust to outliers.
Much research has focused on handling rowwise 
outliers, i.e.
rows that deviate from the majority of the rows in 
the data (for instance, they might belong to a 
different population).
In recent years also cellwise outliers are receiving
attention.
These are suspicious cells (entries) that can occur
anywhere in the table. 
Even a relatively small proportion of outlying cells 
can contaminate over half the rows, which causes 
rowwise robust methods to break down. 
In this paper a new PCA method is constructed which 
combines the strengths of two existing robust methods 
in order to be robust against both cellwise and 
rowwise outliers. 
At the same time, the algorithm can cope with missing 
values. 
As of yet it is the only PCA method that can deal 
with all three problems simultaneously. 
Its name MacroPCA stands for {\bf PCA} allowing 
for {\bf M}issingness {\bf A}nd {\bf C}ellwise \& 
{\bf R}owwise {\bf O}utliers.
Several simulations and real data sets illustrate its 
robustness. 
New residual maps are introduced, which help to 
determine which variables are responsible for the 
outlying behavior. 
The method is well-suited for online process control.
Supplementary material is available online.
\end{abstract}

\vspace{0.3cm}


\section{\normalsize{Introduction}} 
\label{sec:introduction}
 
\begin{figure}[!ht]
\centering
\includegraphics[width=0.42\textwidth]
     {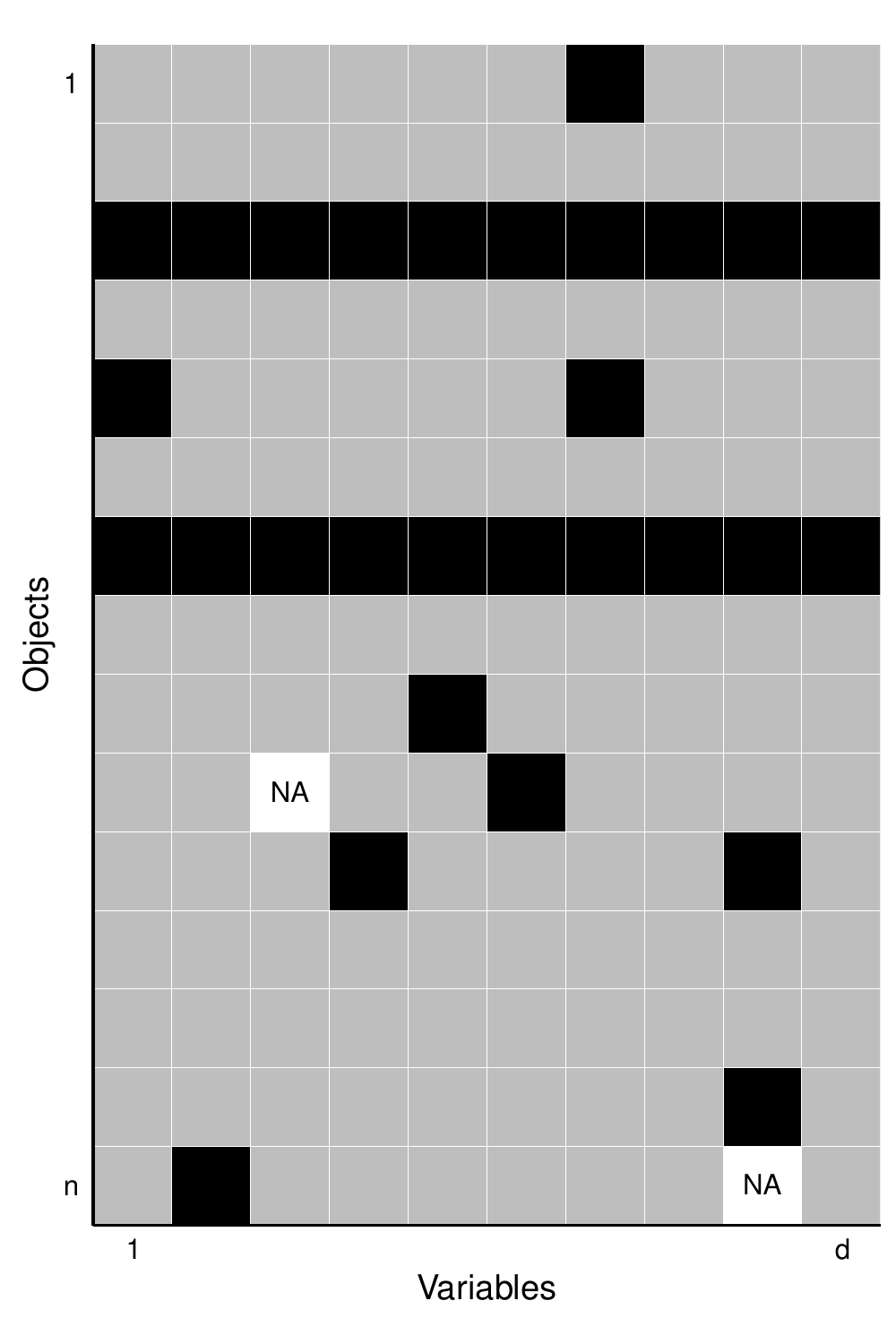}
\caption{Data matrix with missing values and 
         cellwise and rowwise contamination.}
\label{fig:Cellmap_DDC}
\end{figure}

Real data often contain outliers, 
which can create serious problems when analyzing 
it. Many methods have been developed to deal with 
outliers, often by constructing a fit that is robust to 
them and then detecting the outliers by their large
deviation (distance, residual) from that fit. 
For a brief overview of this approach see
\cite{Rousseeuw:WIRE-Anomaly}.
Unfortunately, most robust methods cannot handle data 
with missing values, some rare exceptions being
\cite{Cheng:Missing} and \cite{Danilov:GSE}.
Moreover, they are typically restricted to casewise
outliers, which are cases that deviate from the majority.
We call these {\it rowwise outliers} because multivariate 
data are typically represented by a rectangular matrix 
in which the rows are the cases and the columns are the 
variables (measurements). 
In general, robust methods require that fewer than half
of the rows are outlying, see e.g. 
\cite{Lopuhaa:BDP}.
However, recently a different type of outliers, called
{\it cellwise outliers}, have received much attention 
\citep{Alqallaf:Propag,VanAelst:HSD,Agostinelli:Cellwise}.
These are suspicious cells (entries) that can occur 
anywhere in the data matrix.
Figure \ref{fig:Cellmap_DDC} illustrates the difference
between these types of outliers.
The regular cells are shown in gray, whereas black means
outlying.
Rows 3 and 7 are rowwise outliers, and the other rows
contain a fairly small percentage of cellwise outliers.
As in this example, a small proportion of 
outlying cells can contaminate over half the rows, which 
causes most methods to break down. 
This effect is at its worst when the dimension
(the number of columns) is high. 

In high-dimensional situations, which are becoming
increasingly common, one often applies principal 
component analysis (PCA) to reduce the dimension.
However, the classical PCA (CPCA) method is not robust 
to either rowwise or cellwise outliers. 
Robust PCA methods that can deal with rowwise outliers 
include \citet{Croux:Proj}, \citet{Hubert:RAPCA}, 
\citet{Locantore:Funcdata}, \citet{Maronna:ORreg} and the 
ROBPCA method \citep{Hubert:ROBPCA}. The latter method 
combines projection pursuit ideas with robust covariance 
estimation.

In order to deal with missing values,
\cite{Nelson:miss} and \cite{Kiers:WLS} developed
the {\it iterative classical PCA algorithm} (ICPCA), 
see \citet{Walczak:TutorialI} for a tutorial.
The ICPCA follows the spirit of the EM algorithm.
It starts by replacing the missing values by 
initial estimates such as the columnwise means. 
Then it iteratively fits a CPCA, yielding scores
that are transformed back to the original space
resulting in new estimates for the missing values,
until convergence.

\citet{Serneels:MPCA} proposed a rowwise 
robust PCA method that can also cope with 
missing values. 
We will call this method MROBPCA (ROBPCA for 
missing values) as its key idea is to combine the 
ICPCA and ROBPCA methods. 
MROBPCA starts by imputing the NA's by robust 
initial estimates. The main difference with the 
ICPCA algorithm is that in each iteration the PCA 
model is fit by ROBPCA, which yields different
imputations and flags rowwise outliers.

As of yet there are no PCA methods that can deal 
with cellwise outliers in combination with
rowwise outliers and NA's.
This paper aims to fill that gap by constructing a
new method called MacroPCA, where `Macro' stands
for {\bf M}issingness {\bf A}nd {\bf C}ellwise 
and {\bf R}owwise {\bf O}utliers.
It starts by applying a multivariate method called
DetectDeviatingCells \citep{Rousseeuw:DDC} for 
detecting cellwise outliers, which provides initial 
imputations for the outlying cells and the NA's
as well as an initial measure of rowwise outlyingness.
In the next steps MacroPCA combines ICPCA and
ROBPCA to protect against rowwise outliers and to
create improved imputations of the outlying cells 
and missing values. 
MacroPCA also provides graphical displays to 
visualize the different types of outliers.
R code for MacroPCA is publicly available
(Section \ref{sec:software}).

\section{The MacroPCA algorithm}
\label{sec:MacroPCA}

\subsection{Model}
\label{subsec:Model}

The data matrix is denoted as $\bX_{n,d}$ in which 
the subscripts are the number of rows (cases) $n$ 
and the number of columns (variables) $d$ .
In the absence of outliers and missing values the
goal is to represent the data in a lower dimensional 
space, i.e.\
\begin{equation} \label{eq:pcamodel}
	\bX_{n,d} = \bone_n \bmu_d' +
	\bmT_{n,k} (\bmP_{d,k})' + \bmE_{n,d} 
\end{equation}
with $\bone_n$ the column vector with all $n$ 
components equal to 1, $\bmu_d$ the 
$d$-variate column vector of location,  
$\bmT_{n,k}$ the $n \times k$ score matrix,
$\bmP_{d,k}$ the $d \times k$ loadings matrix 
whose columns span the PCA subspace, and
$\bmE_{n,d}$ the $n \times d$ error matrix. 
The reduced dimension $k$ can vary from 
1 to $d$ but we assume that $k$ is low. 
The $\bmu_d$\,, $\bmT_{n,k}$\, and $\bmP_{d,k}$
are unknown, and estimates of them will be
denoted by $\bm_d$\,, $\bT_{n,k}$ and 
$\bP_{d,k}$\,.

Several realities complicate this simple model.
First, the data matrix may not be fully observed, 
i.e., some cells $x_{ij}$ may be missing.
Here we assume that they are 
\textit{missing at random} (MAR), 
meaning that the missingness of a cell is
unrelated to the value the cell would have had,
but may be related to the values of other
cells in the same row; see, e.g., 
\cite{Schafer:missing}.
This is the typical assumption underlying
EM-based methods such as ICPCA and MROBPCA
that are incorporated in our proposal.

Secondly, the data may contain rowwise outliers,
e.g. cases from a different population.
The existing rowwise robust methods require 
that fewer than half of the rows are outlying, 
so we make the same assumption here. 

Thirdly, cellwise outliers may occur
as described in the introduction. 
The outlying cells may be imprecise, incorrect or 
just unusual. 
Outlying cells do not necessarily stand out in their 
column because the correlations between the columns
matter as well, so these cells may not be detectable by 
simple univariate outlier detection methods.
There can be many cellwise outliers, and
in fact each row may contain one or
more outlying cells. 

\subsection{Dealing with missing values 
            and cellwise and rowwise outliers}
\label{subsec:Algorithm}

We propose the MacroPCA algorithm for analyzing data 
that may contain one or more of the following issues:
missing values, cellwise outliers, and rowwise outliers.
Throughout the algorithm we will use the following
two notations:
\begin{itemize}
\item the {\it NA-imputed matrix} $\bcX_{n,d}$ only 
      imputes the missing values of $\bX_{n,d}$\;;
\item the {\it cell-imputed matrix} $\bfX_{n,d}$ has 
      imputed values for the outlying cells that do not 
			belong to outlying rows, and for all missing 
			values.
\end{itemize}
Both of these matrices still have $n$ rows.
Neither is intended to simply replace the true data 
matrix $\bX_{n,d}$\;.
Note that $\bfX_{n,d}$ does not try to impute outlying 
cells inside outlying rows, which would mask these rows 
in subsequent computations.

Since we do not know in advance which cells and 
rows are outlying, the set of flagged cellwise 
and rowwise outliers (and hence $\bcX_{n,d}$ and
$\bfX_{n,d}$) will be updated in the course of
the algorithm. 

The first part of MacroPCA is the 
DetectDeviatingCells (DDC) algorithm. 
The description of this method can be found in
\cite{Rousseeuw:DDC} and in 
Section 1 of the Supplementary Material. 
The main purpose of the DDC method is to detect 
cellwise outliers.  
DDC outputs their positions $I_{c,DDC}$ as
well as imputations for these outlying cells 
and any missing values.
It also yields an initial outlyingness 
measure on the rows, which is however not 
guaranteed to flag all outlying rows.
The set of flagged rows $I_{r,DDC}$ will be 
improved in later steps. 

The second part of MacroPCA constructs principal 
components along the lines of the ICPCA 
algorithm but employing a version of ROBPCA 
\citep{Hubert:ROBPCA} to fit subspaces. 
It consists of the following steps, with all
notations listed in Section 2 
of the Supplementary Material.
\begin{enumerate}
\item {\bf Projection pursuit.} The goal of this
step is to provide an initial indication of which
rows are the least outlying. 
For this ROBPCA starts by 
identifying the $h < n$ least outlying rows by a 
projection pursuit procedure. 
We write $0.5 \leqslant \alpha=h/n < 1$.
This means that we can withstand up to a fraction 
$1-\alpha$ of outlying rows. 
To be on the safe side the default is
$\alpha=0.5$\,.
	
However, due to cellwise outliers there may be far
fewer than $h$ uncontaminated rows, so we cannot 
apply this step to the original data $\bX_{n,d}$. 
We also cannot use the entire imputed matrix 
$\btX_{n,d}$ obtained 
from DDC in which all outlying cells are imputed, even 
those in potentially outlying rows, as this could mask 
outlying rows. 
Instead we use the cell-imputed matrix 
$\bfX_{n,d}^{(0)}$ defined as follows: 
\begin{enumerate}
\item In all rows flagged as outlying we keep the
   original data values. Only the missing values in
	 these rows are replaced by the values imputed by
	 DDC. More precisely, for all $i$ in $I_{r,DDC}$
	 we set $\bfx_i^{(0)} = \bcx_i^{(0)}$. 
\item In the $h$ unflagged rows with the fewest cells 
flagged by DDC we impute those cells, 
i.e. $\bfx_i^{(0)} = \btx_i$. 
\end{enumerate}
As in ROBPCA the outlyingness of a point 
$\bfx_i^{(0)}$ is then computed as
\begin{equation}
	\text{outl}(\bfx_i^{(0)}) = \max_{\bv \in B} 
	\frac{|\bv'\bfx_i^{(0)} - \lmcd(\bv'\bfx_j^{(0)})|}
	{\smcd(\bv'\bfx_j^{(0)})} \label{outlo} \ \ ,
\end{equation}
where $\lmcd(\bv'\bfx_j^{(0)})$ and 
$\smcd(\bv'\bfx_j^{(0)})$ are the univariate MCD location 
and scale estimators \citep{Rousseeuw:RobReg} of 
$\{\bv'\bfx_1^{(0)},\ldots,\bv'\bfx_n^{(0)}\}$ . 
The set $B$ contains 250 directions through two data 
points (or all of them if there are fewer than 250).
Finally, the indices of the $h$ rows $\bfx_i^{(0)}$ 
with the lowest outlyingness and not belonging to 
$I_{r,DDC}$ are stored in the set $H_0$\,.

\item {\bf Subspace dimension.}
Here we choose the number of principal components.
For this we build a new cell-imputed matrix 
$\bfX^{(1)}_{n,d}$ which imputes the outlying cells in
the rows of $H_0$ and imputes the NA's in all rows. 
This means that $\bfx_i^{(1)} = \btx_i$ for 
$i \in H_0$\,, and $\bfx_i^{(1)} = \bcx_i^{(0)}$ if 
$i \notin H_0$. 
Then we apply classical PCA to the $\bfx_i^{(1)}$ with 
$i \in H_0$.  
Their mean $\bm_{d}^{(1)}$ is an estimate of the center, 
whereas the spectral decomposition of their covariance 
matrix yields a loading matrix $\bP_{d,d}^{(1)}$ and a 
diagonal matrix $\bL_{d,d}^{(1)}$ with the eigenvalues 
sorted from largest to smallest. 
These eigenvalues can be used to construct a screeplot 
from which an appropriate dimension $k$ of the subspace 
can be derived. 
Alternatively, one can retain a certain cumulative 
proportion of explained variance, such as 80\%.
The maximal number of principal components that MacroPCA
will consider is the tuning constant $k_{\max}$ which is
set to 10 by default.  

\item {\bf Iterative subspace estimation.}
This step aims to estimate the $k$-dimensional 
subspace fitting the data.
As in ICPCA this requires iteration, for $s \gs 2$:  
\begin{enumerate}
\item The scores matrix in \eqref{eq:pcamodel} based 
on the cell-imputed cases is computed as 
$\bfT_{n,k}^{(s-1)} = (\bfX_{n,d}^{(s-1)} - 
\boldsymbol 1_n 
(\bm_{d}^{(s-1)})') \bP_{d,k}^{(s-1)}$\;.
The predicted data values are set to 
$\bhX_{n,d}^{(s)} = \boldsymbol 1_n (\bm_d^{(s-1)})' 
+ \bfT_{n,k}^{(s-1)} (\bP_{d,k}^{(s-1)})'$\;.
We then update the imputed matrices to 
$\bcX_{n,d}^{(s)}$ and $\bfX_{n,d}^{(s)}$ by replacing 
the appropriate cells by the corresponding cells of 
$\bhX_{n,d}^{(s)}$. 
That is, for $\bcX_{n,d}^{(s)}$ we update all the 
imputations of missing cells, whereas for  
$\bfX_{n,d}^{(s)}$ we update the imputations of the 
outlying cells in rows of $H_0$ as well as the 
NA's in all rows.
\item The PCA model is re-estimated	by applying 
classical PCA to the $\bfx_i^{(s)}$ with $i \in H_0$.
This yields a new estimate $\bm_{d}^{(s)}$
as well as an updated loading matrix 
$\bP_{d,k}^{(s)}$\;. 
\end{enumerate}
The iterations are repeated until $s=20$ or until 
convergence is reached, i.e.\ when 
the maximal angle between a vector in the new subspace 
and the vector most parallel to it in the previous 
subspace is below some tolerance
(by default 0.005).
Following~\citet{Krzanowski:GroupPC} this angle is 
computed as $arccos(\sqrt{\delta_k})$
where $\delta_k$ is the smallest eigenvalue of 
$(\bP_{d,k}^{(s)})' \bP^{(s-1)}_{d,k} 
(\bP^{(s-1)}_{d,k})' \bP_{d,k}^{(s)}$\;.

After all iterations we have the cell-imputed
matrix $\bfX_{n,d}^{(s)}$ as well as the estimated 
center $\bm_{d}^{(s)}$ and the 
loading matrix $\bP_{d,k}^{(s)}$\,. 

\item {\bf Reweighting.} 
In robust statistics one often follows an initial
estimate by a reweighting step in order to improve
the statistical efficiency at a low computational
cost, see e.g. 
\citep{Rousseeuw:RobReg,Engelen:PCA}. 
Here we use the orthogonal distance of each 
$\bfx_i^{(s)}$ to the current PCA subspace: 
\begin{equation*}
  \fod_i = \|\bfx_i^{(s)} - \bfhx_i^{(s)}\| =  
	\| \bfx_i^{(s)}- (\bm_d^{(s)} + 
	(\bfx^{(s)}_i- \bm_d^{(s)})\bP_{d,k}^{(s)}
	(\bP_{d,k}^{(s)})') \| \ .
\end{equation*} 
The orthogonal distances to the power 2/3 are roughly 
Gaussian except for the outliers \citep{Hubert:ROBPCA}, 
so we compute the cutoff value 
\begin{equation}\label{eq:cutoffOD}
  c_{\od} := 
	 \left(\lmcd(\{\fod_j^{2/3}\})+
	  \smcd(\{\fod_j^{2/3}\}) 
	 \, \Phi^{-1}(0.99) \right)^{3/2} \;\;. 
\end{equation} 
All cases for which $\fod_i \ls c_{\od}$ 
are considered non-outlying with respect to the 
PCA subspace, and their indices are stored in a 
set $H^*$. As before, any $i \in I_{r,DDC}$ is 
removed from $H^*$. The cases not in $H^*$ are 
flagged as rowwise outliers.
The final cell-imputed matrix $\bfX_{n,d}$ 
is given by 
$\fx_{i,j} = \hfx_{i,j}^{(s)}$ if 
$i \in H^*$ and $j \in I_{c,DDC}$ and
$\fx_{i,j} = \cx_{i,j}$ otherwise.
Applying classical PCA to the $n^*$ rows 
$\bfx_i$ in $H^*$ yields a new center $\bm_d^*$ 
and a new loading matrix $\bP_{d,k}^*$\;.

\item {\bf DetMCD.} Now we want to estimate
a robust basis of the estimated subspace.
The columns of $\bP_{d,k}^*$ from step 4 need not 
be robust, because some of the $n^*$ rows in 
$H^*$ might be outlying inside the subspace. 
These so-called 
good leverage points do not harm the 
estimation of the PCA subspace but they can still 
affect the estimated eigenvectors and eigenvalues,
as illustrated by a toy example in Section 
\ref{A:toy} of the Appendix.
In this step we first project the $n^*$ points 
of $H^*$ onto the subspace, yielding
\begin{equation*}
\bfT_{n^*,k} = \left(\bfX_{n^*,d} - 
  \boldsymbol 1_{n^*} \bm_d^{*'} \right)
	\bP_{d,k}^* \ \ .
\end{equation*}
Next, the center and scatter matrix of the scores 
$\bfT_{n^*,k}$ are estimated by the DetMCD method 
of \citet{Hubert:DetMCD}. 
This is a fast, robust and deterministic algorithm 
for multivariate location and scatter, yielding 
$\bm_k^{\mcd}$ and $\bS_{k,k}^{\mcd}$.
Its computation is feasible because the dimension 
$k$ of the subspace is quite low.
The spectral decomposition of $\bS_{k,k}^{\mcd}$
yields a loading matrix $\bP_{k,k}^{\mcd}$
and eigenvalues $\hlam_j$ for $j=1,\ldots,k$\;.
We set the final center to 
$\bm_d = \bm_d^{*} + 
\bP_{d,k}^{*}\bm_k^{\mcd}$ 
and the final loadings to 
$\bP_{d,k} = \bP_{d,k}^*\bP_{k,k}^{\mcd}$. 

\item{\bf Scores, predicted values and residuals.}
We now provide the final output.
We compute the scores of $\bcX_{n,d}$
as $\bcT_{n,k} = (\bcX_{n,d} - 
\boldsymbol 1_{n}\bm_d')\bP_{d,k}$
and the predictions of $\bcX_{n,d}$ 
as $\bchX_{n,d}  = \boldsymbol 1_{n} \bm_d' +
\bcT_{n,k} (\bP_{d,k})'$\;.
(The formulas for $\bfT_{n,k}$ and $\bfhX_{n,d}$
are analogous.)
This yields the difference matrix  
$\bcX_{n,d}-\bchX_{n,d}$ which we then 
robustly scale by column, yielding the
final standardized residual matrix $\bR_{n,d}$\,.
The orthogonal distance of $\bcx_i$ to the PCA 
subspace is given by
\begin{equation}\label{eq:od}
\cod_i = \| \bcx_i - \bchx_i \| \;.
\end{equation}
\end{enumerate}
See Section \ref{sec:software} for the R code
carrying out MacroPCA.

\begin{figure}[!ht]
\centering
\begin{tabular}{cc}
\includegraphics[width=0.45\textwidth]
  {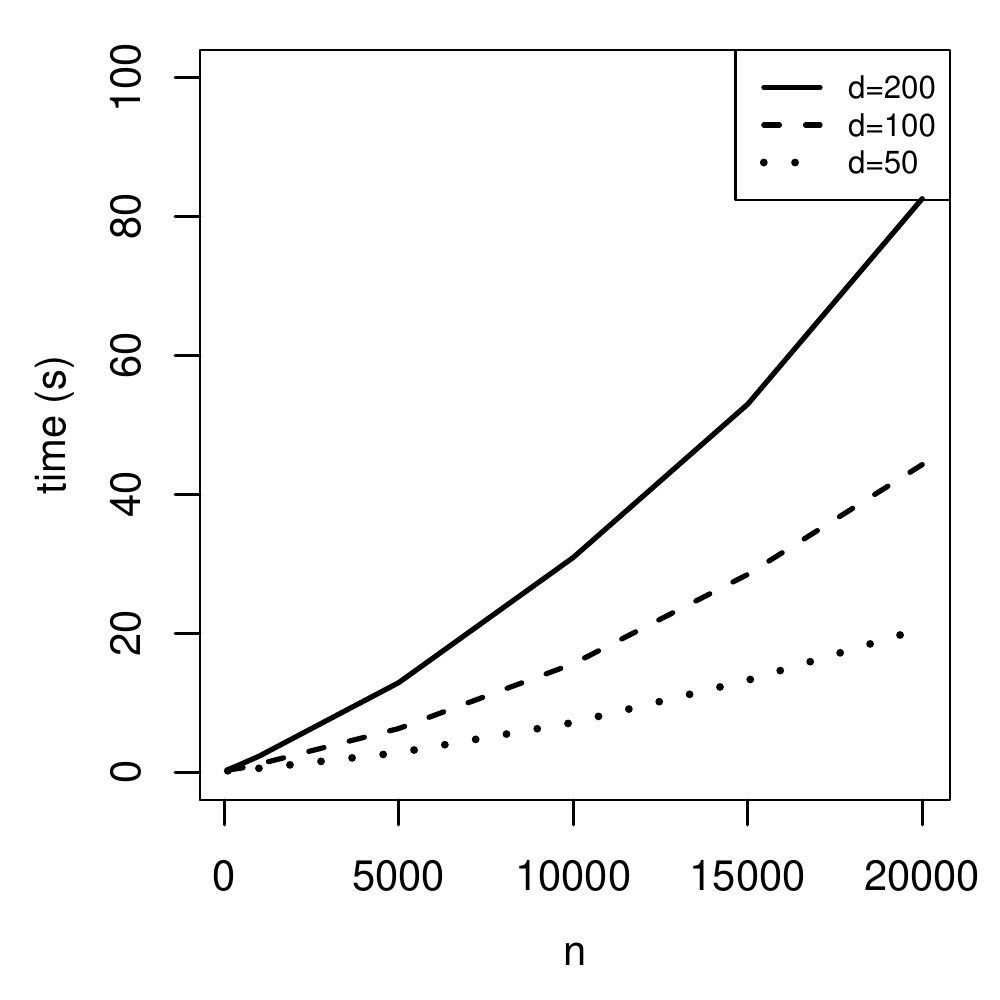} & 
\includegraphics[width=0.45\textwidth]
  {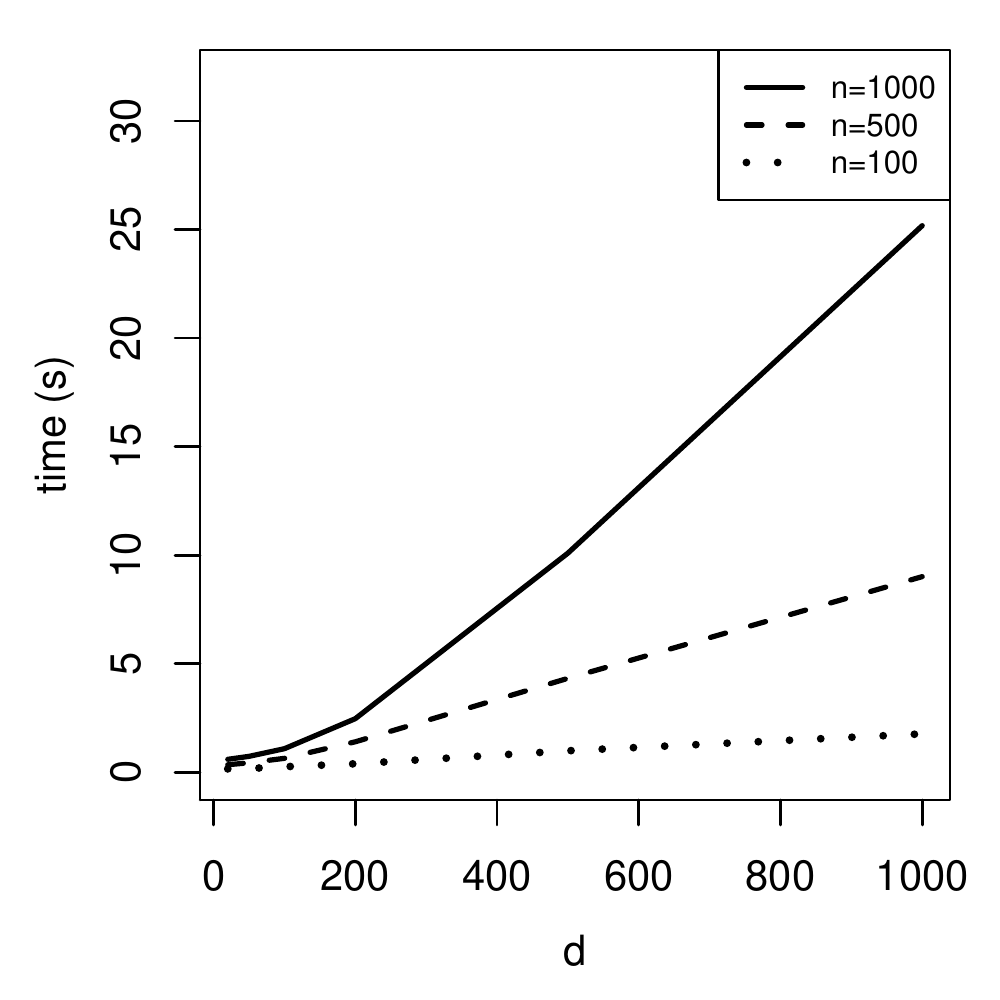}
\vspace{-0.4cm}	
\end{tabular}
\caption{Computation times of MacroPCA in
  seconds on Intel i7-4800MQ at 2.70 GHz,
	as a function of the number of
	cases $n$ (left) and of the dimension
	$d$ (right).}
	\label{fig:times}
\end{figure}

MacroPCA can be carried out 
in $O(nd(\min(n,d) +\log(n) + \log(d)))$
time (see\linebreak 
Section \ref{A:complexity} of the Appendix)
which is not much more than the 
complexity\linebreak  
$O(nd\min(n,d))$ of classical PCA.
Figure \ref{fig:times} shows times as a 
function of $n$ and $d$ indicating that
MacroPCA is quite fast.
The fraction of NA's in the data had no
substantial effect on the computation
time, as seen in Figure 
\ref{fig:timesNAs}
in Section \ref{A:complexity}.

Note that PCA loadings are highly influenced by 
the variables with the largest variability. 
For this the MacroPCA code provides the option to 
divide each variable by a robust scale. This does 
not increase the computational complexity.

\section{Outlier detection}
\label{sec:detection}

MacroPCA provides several tools for outlier detection. 
We illustrate them on a dataset collected by
\cite{Alfons:robustHD} from the website of the 
Top Gear car magazine. 
It contains data on 297 cars, with 11 continuous 
variables.
Five of these variables (price, displacement, BHP, 
torque, top speed) are highly skewed, and were
logarithmically transformed. 
The dataset contains 95 missing cells, which is only 
2.9\% of the $297 \times 11 = 3267$ cells. 
We retained two principal components ($k=2$).

\begin{figure}[ht!]
\centering
\includegraphics[width=1.0\textwidth]
	{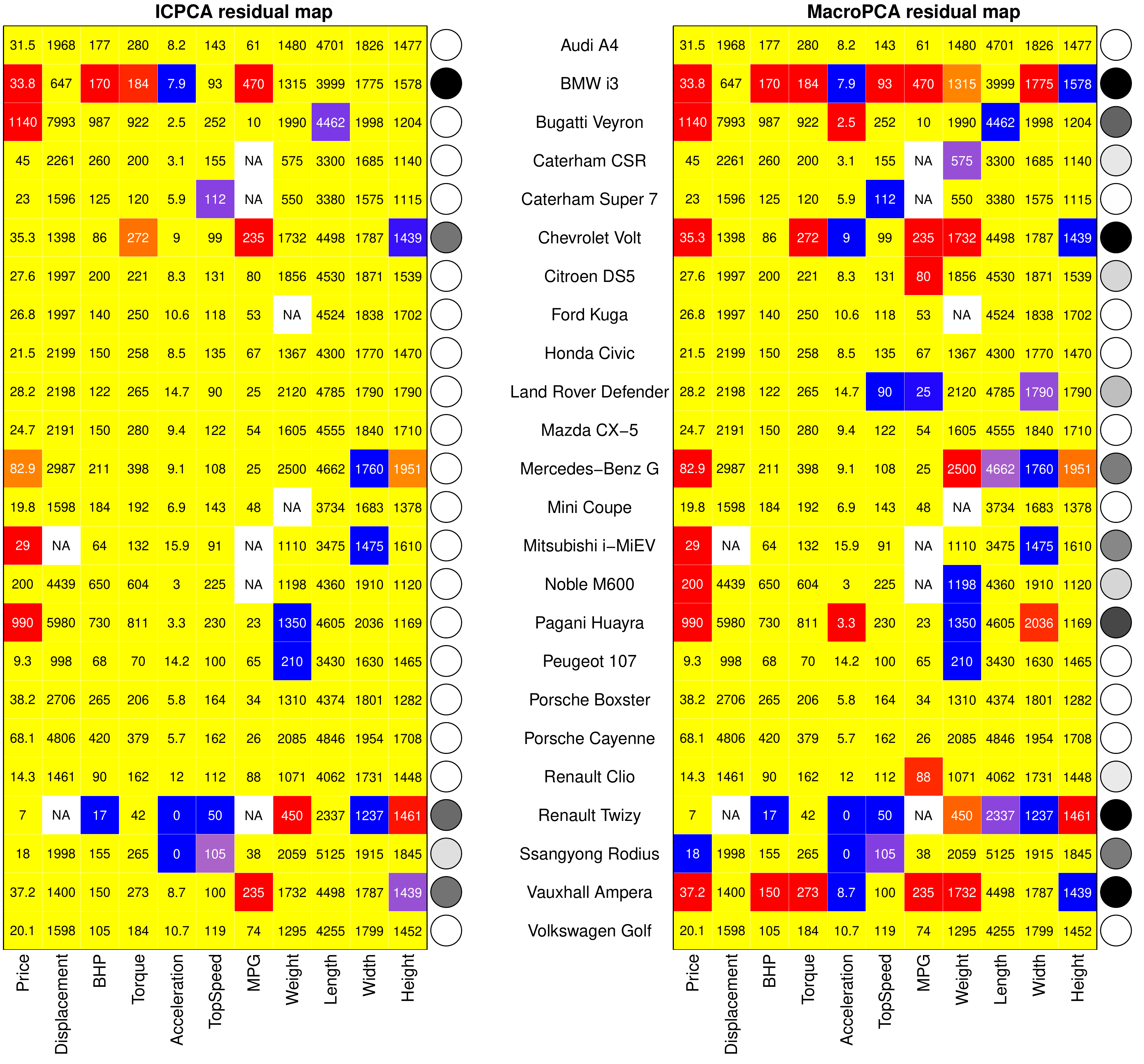}
\vspace{-1.0cm}
\caption{Residual map of selected rows from Top Gear 
  data: (left) when using ICPCA; (right) when using 
	MacroPCA. The numbers shown in the cells are
	the original data values (with price in units
	of 1000 UK Pounds).}
\label{fig:Cars1}
\end{figure}

The right hand panel of Figure~\ref{fig:Cars1} shows 
the results of MacroPCA by a modification of the 
cell map introduced by \citet{Rousseeuw:DDC}.
The computations were performed on all 297 cars, but
in order to make the map fit on a page it only shows 
24 cars, including some of the more eventful cases.
The color of the cells stems from the standardized 
residual matrix $\bR_{n,d}$ obtained by MacroPCA. 
Cells with $|r_{ij}| \ls \sqrt{\chi^2_{1,0.99}} = 2.57$ 
are considered regular and colored yellow in the 
residual map, whereas the missing values are white.
Outlying residuals receive a color which ranges from light 
orange to red when $r_{ij} > 2.57$ and from light purple
to dark blue when $r_{ij} < -2.57$\;. 
So a dark red cell indicates that its observed value is 
much higher than its fitted value, while a dark blue 
cell means the opposite. 

To the right of each row in the map is a circle 
whose color varies from white to black according 
to the orthogonal distance $\cod_i$ given by
\eqref{eq:od} compared to the cutoff 
\eqref{eq:cutoffOD}.
Cases with $\cod_i \ls c_{\od}$ lie close to the PCA 
subspace and receive a white circle. 
The others are given darker shades of gray up to black
according to their $\cod_i$\;. 
 
On these data we also ran the ICPCA method, which handles
missing values in classical PCA. 
It differs from MacroPCA in some important ways: the
initial imputations are by nonrobust column means, the
iterations carry out CPCA and do not exclude outlying 
rows, and the residuals are standardized by the 
nonrobust standard deviation.
By itself ICPCA does not provide a residual map, but 
we can construct one anyway by plotting the nonrobust
standardized residuals with the same color scheme,
yielding the left panel of Figure~\ref{fig:Cars1}. 

The ICPCA algorithm finds high orthogonal distances
(dark circles)
for the BMW i3, the Chevrolet Volt, the Renault Twizzy
and the Vauxhall Ampera.
These are hybrid or purely electrical cars with a high 
or missing MPG (miles per gallon).
Note that the Ssangyong Rodius and Renault Twizzy get
blue cells for their acceleration time of zero seconds, 
which is physically impossible. 

On this dataset the ICPCA algorithm provides decent results
because the total number of outliers is small compared
to the size of the data, and indeed the residual map of 
all 297 cars was mostly yellow.
But MacroPCA (right panel) detects more deviating behavior. 
The orthogonal distance of the hybrid Citroen DS5 and the 
electrical Mitsubishi i-MiEV are now on the high side,
and the method flags the Bugatti Veyron and Pagani Huayra 
supercars as well as the Land Rover Defender and 
Mercedes-Benz G all-terrain vehicles.
It also flags more cells, giving a more complete
picture of the special characteristics of some cars.

\begin{figure}[!ht]
\centering
\begin{tabular}{cc}
\includegraphics[width=0.48\textwidth]
	{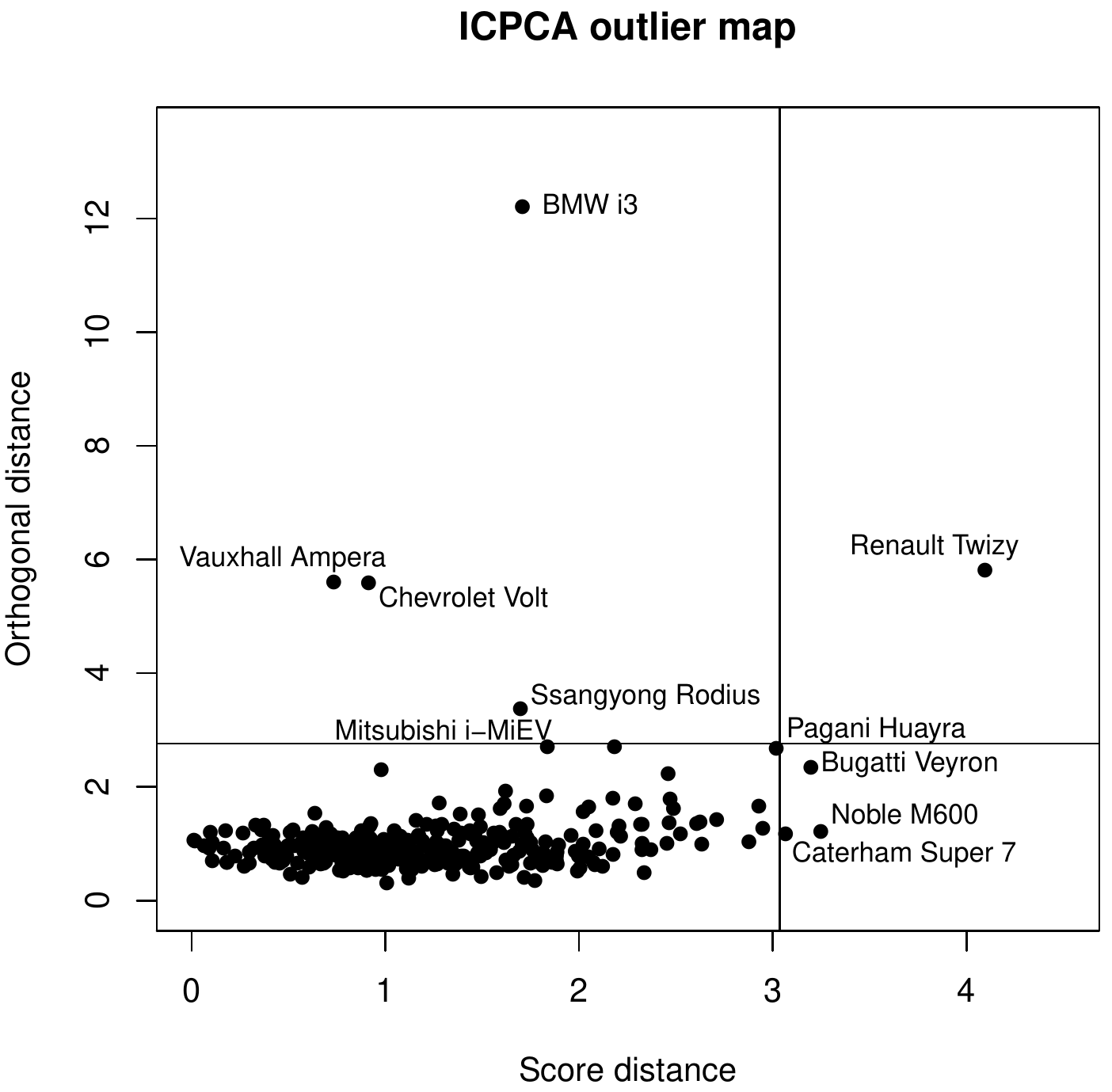} & 
\includegraphics[width=0.48\textwidth]
	{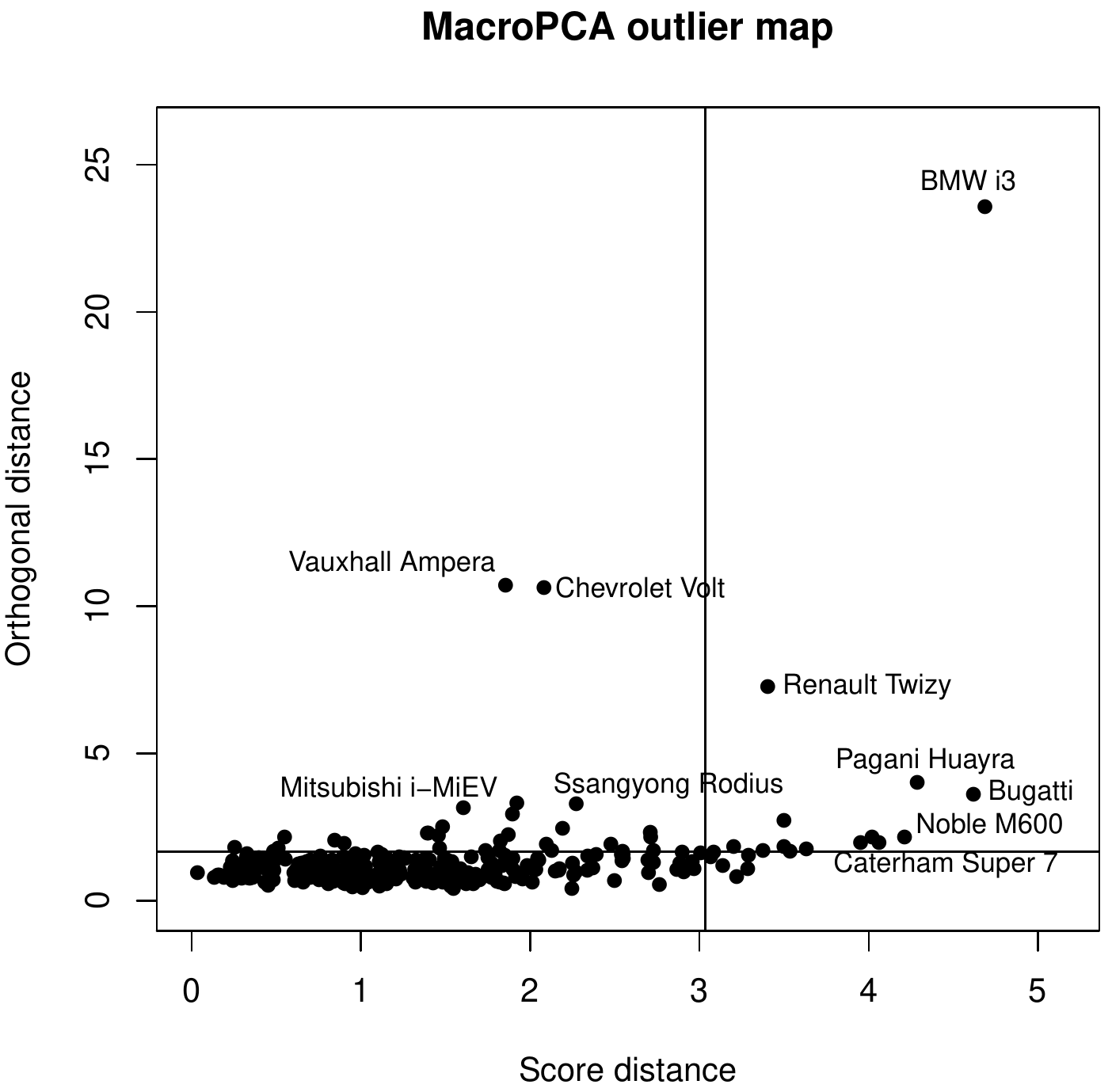} 
\end{tabular}
\vspace{-0.7cm}
\caption{Outlier map of Top Gear data: (left) when 
         using ICPCA; (right) when using MacroPCA.}
\label{fig:Cars2}
\end{figure}

We can also compute the {\it score distance} of each
case, which is the robustified Mahalanobis distance of
its projection on the PCA subspace among all such
projected points.
It is easily computed as
\begin{equation}
\csd_i = \sqrt{\sum_{j=1}^k (\ct_{ij})^2/\hlam_j}   
\label{eq:sd}
\end{equation}
where $\ct_{ij}$ are the scores and $\hlam_j$ the 
eigenvalues obtained by MacroPCA.
This allows us to construct a PCA outlier map of cases 
as introduced in ~\citet{Hubert:ROBPCA}, which plots
the orthogonal distances $\cod_i$ on the vertical axis 
versus the score distances $\csd_i$\;.
The MacroPCA outlier map of these data is the right
panel of Figure~\ref{fig:Cars2}.
The vertical line indicates the cutoff
$c_{\sd} = \sqrt{\chi^2_{k,0.99}}$ 
and the horizontal line is the cutoff $c_{\od}$\,.

Regular cases are those with a small 
$\csd_i \ls c_{\sd}$ and a small $\cod_i \ls c_{\od}$\;.
Cases with large $\csd_i$ and small $\cod_i$ are called
good leverage points. 
The cases with large $\cod_i$ can be divided into 
orthogonal outliers (when their $\csd_i$ is small) and 
bad leverage points (when their $\csd_i$ is large too). 
We see several orthogonal outliers such as the Vauxhall 
Ampera as well as some bad leverage points, especially 
the BMW i3.
There are also some good leverage points. 

The left panel displays the outlier map for ICPCA.
It flags the BMW i3 as an orthogonal outlier. 
This behavior is typical because a bad leverage point
will attract the fit of classical methods, making it 
appear less special.
For the same reason ICPCA considers some of the good 
leverage points as regular cases.
That the ICPCA outlier map is still able to flag some
outliers is due to the fact that this dataset
only has a small percentage of outlying rows.

\section{Online data analysis}
\label{sec:predict}

Applying MacroPCA to a data set $\bX_{n,d}$ yields a
PCA fit.
Now suppose that a new case (row) $\bx$ comes in, and 
we would like to impute its missing values,
detect its outlying cells and impute them, estimate 
its scores, and find out whether it is a rowwise outlier.
We could of course append $\bx$ to $\bX_{n,d}$ and rerun 
MacroPCA, but that would be very inefficient.

Instead we propose a method to analyze $\bx$ using
only the output of MacroPCA on the initial 
set $\bX_{n,d}$\;.
This can be done quite fast, which makes the procedure 
suitable for online process control. 
For outlier-free data with NA's this was
studied by \cite{Nelson:miss} and
\cite{Walczak:TutorialI}. 
\cite{FolchFortuny:PCAmissing} call this model
exploitation, as opposed to model building
(fitting a PCA model).
Our procedure consists of two stages, along the 
lines of MacroPCA.  
\begin{enumerate}[label={\arabic*.}]
\item {\bf DDCpredict} is a new function
which only uses $\bx$ and the output of DDC on the 
initial data $\bX_{n,d}$\;. 
First the entries of $\bx$ are standardized using 
the robust location and scale estimates from DDC. 
Then all $x_j$ with 
$|x_j| > \sqrt{\chi^2_{1,0.99}} = 2.57$
are replaced by NA's.
Next all NA's are estimated 
as in DDC making use of the pre-built 
coefficients $b_{jh}$ and weights $w_{jh}$\;.
Also the deshrinkage step uses the original
robust slopes. 
The \textit{DDCpredict} stage yields the imputed 
vector $\btx^{(0)}$ and the standardized residual 
of each cell $x_j$.  

\item {\bf MacroPCApredict} improves on the initial 
imputation $\btx^{(0)}$\;.
The improvements are based solely on the $\bm_d$ 
and $\bP_{d,k}$ that were obtained by MacroPCA
on the original data $\bX_{n,d}$\;.
Step $s \gs 1$ is of the following form:
\begin{enumerate}
\item Project the imputed case $\btx^{(s-1)}$ 
	on the MacroPCA subspace to obtain its scores 
	vector
	$\bt^{(s)} = (\bP_{d,k})'(\btx^{(s-1)} - \bm_d)$;
\item transform the scores to the original 
  space, yielding 
	$\hat{\bx}^{(s)} = \bm_d + \bP_{d,k} \bt^{(s)}$\;;
\item Reimpute the outlying cells and missing values 
  of $\bx$ by the corresponding values of 
	$\hat{\bx}^{(s)}$, yielding $\btx^{(s)}$\,.
\end{enumerate}
These steps are iterated until convergence 
(when the new imputed values are within a tolerance
of the old ones) or
the maximal number of steps (by default 20) is 
reached.
We denote the final $\btx^{(s)}$ as $\btx$.

Next we create $\bcx$ by replacing
the missing values in $\bx$ by the corresponding 
cells in $\btx$.
We then compute the orthogonal distance
$\OD(\bcx)$ and the score distance $\SD(\bcx)$.
If $\OD(\bcx) > c_{\od}$ the new case $\bx$ is
flagged as an orthogonal outlier.
Finally the cell residuals $\cx_j - \chx_j$ 
are standardized as in the last step of MacroPCA, 
and used to flag outlying cells in $\bx$.
\end{enumerate}

\begin{figure}[!ht]
\centering
\includegraphics[width=1.0\textwidth]
	{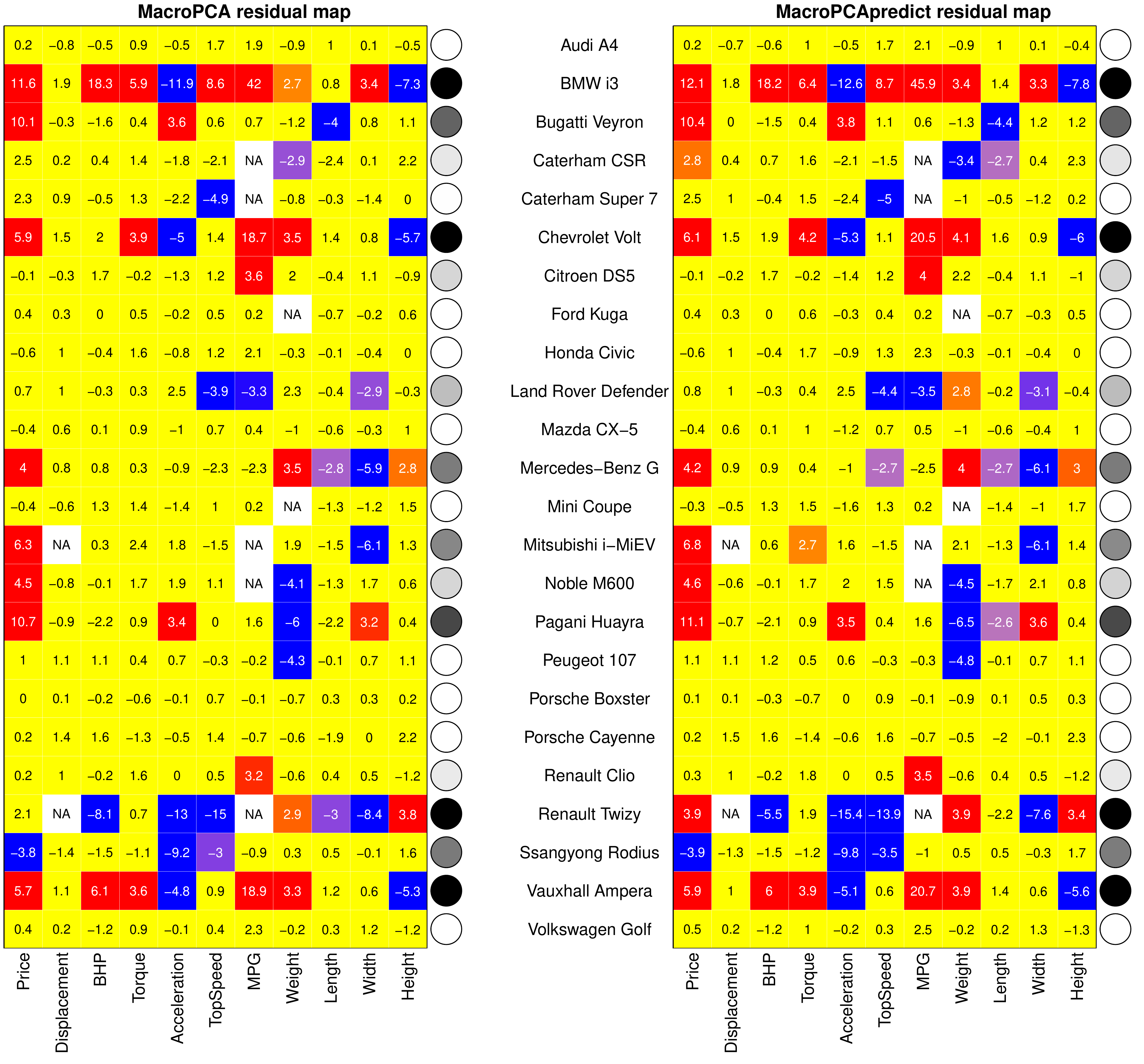}
\vspace{-1.0cm}
\caption{Top Gear data set: residual maps obtained by 
  (left) including and (right) excluding these 24 
	cars when fitting the PCA model.}
\label{fig:Cars3}
\end{figure}

To illustrate this prediction 
procedure we re-analyze the Top Gear data set. 
We exclude the 24 cars shown in the residual map 
of Figure~\ref{fig:Cars1} 
and build the MacroPCA model on the remaining data. 
This model was then provided to analyze the 24 
selected cars as `new' data.
Figure~\ref{fig:Cars3} shows the result. 
As before the cells are colored according to their 
standardized residual, and the circles on 
the right are filled according to their $\cod$. 
The left panel is the MacroPCA residual map shown 
in Figure~\ref{fig:Cars1}, which was obtained by 
applying MacroPCA to the entire data set. 
The right panel shows the result of analyzing these
24 cases using the fit obtained without them.
The residual maps are quite similar. 
Note that each cell now shows its standardized 
residual (instead of its data value as in
Figure~\ref{fig:Cars1}), making it
easier to see the differences. 

\section{Simulations}
\label{sec:Simulations}

We have compared the performance of ICPCA, MROBPCA 
and MacroPCA in an extensive simulation study.
Several contamination models were considered with
missing values, cellwise outliers, rowwise outliers, 
and combinations of them. Only a few of the results 
are reported here since the others yielded similar 
conclusions. 

The clean data $\bX_{n,d}^0$ are generated from a
multivariate Gaussian with $\bmu = \bzero$ and two 
types of covariance matrices $\bSigma_{d,d}$. 
The first one is based on the structured 
correlation matrix called A09 where each off-diagonal 
entry is $\rho_{i,j} = \left(-0.9\right)^{|i-j|}$. 
The second type of covariance matrix is based on the 
random correlation matrices of 
\cite{Agostinelli:Cellwise} and will be called ALYZ. 
These correlation matrices are turned into covariance 
matrices with other eigenvalues. 
More specifically, the diagonal elements of the 
matrix $\bL_{d,d}$ from the spectral decomposition 
$\bSigma_{d,d}={\bP_{d,d}}\bL_{d,d}{\bP'_{d,d}}$ 
are replaced by the
desired values listed below. 
The specifications of the clean
data are $n=100$, $d=200$, 
$\bL_{d,d} = \text{diag}(30, 25, 20, 15, 10, 5, 
0.098, 0.0975, \ldots, 0.0020, 0.0015)$ and $k=6$ 
(since $\sum_{j=1}^{6} \lambda_j / 
\sum_{j=1}^{200} \lambda_j = 91.5\% $). 
MacroPCA takes less than a second for $n=100$, 
$d=200$ as seen in Figure \ref{fig:times}.

In a first simulation setting, the clean data 
$\bX_{n,d}^0$ are modified by replacing a random 
subset of 5\%, 10\%, ... up to 30\% of the 
$n \times d$ cells with NA's. 
The second simulation setting generates NA's and 
outlying cells by randomly replacing 20\% of the cells 
$x_{ij}$ by missing values and 20\% by the value 
$\gamma\sigma_j$ 
where $\sigma^2_{j}$ is the $j$-th diagonal element of 
$\bSigma_{d,d}$ and $\gamma$ ranges from 0 to 20.  
The third simulation setting generates NA's and 
outlying rows. 
Here 20\% of random cells are replaced by NA's 
and a random subset of 20\% of the rows is replaced 
by rows generated from 
$N(\gamma \bv_{k+1},\bSigma_{d,d})$ where 
$\gamma$ varies from 0 to 50 and $\bv_{k+1}$ is the 
$(k+1)$th eigenvector of $\bSigma_{d,d}$. 
The last simulation setting generates 20\% of NA's, 
together with 10\% of cellwise outliers and 10\% of
rowwise outliers in the same way.  

In each setting we consider the set C consisting 
of the rows $i$
that were not replaced by rowwise outliers, 
with $c := \#C$, and the data matrix
$\bX_{c,d}^0$ consisting of those rows of the
clean data $\bX_{n,d}^0$\,.
As a baseline for the simulation we apply 
classical PCA to $\bX_{c,d}^0$ and denote the
resulting predictions by $\hat{x}_{ij}^C$ for 
$i$ in $C$. 
We then measure the mean squared error (MSE) from 
the baseline:
\begin{equation*} 
	\text{MSE} = \frac{1}{c d}\,
	  \sum_{i \in C} \sum_{j=1}^{d} 
		\left(\hat{x}_{ij} - \hat{x}_{ij}^{C}\right)^2
	\end{equation*} 
where $\hat{x}_{ij}$ is the predicted value for 
$x_{ij}$ obtained by applying the different 
methods to the contaminated data.
The MSE is then averaged over 100 replications. 

\begin{figure}[!ht]
\centering
\vskip2mm
\begin{tabular}{cc}
 \hskip5mm A09, fraction $\varepsilon$ of missing 
  values & \hskip0mm ALYZ, fraction $\varepsilon$ 
	of missing values \\
 \includegraphics[width=0.45\textwidth]
   {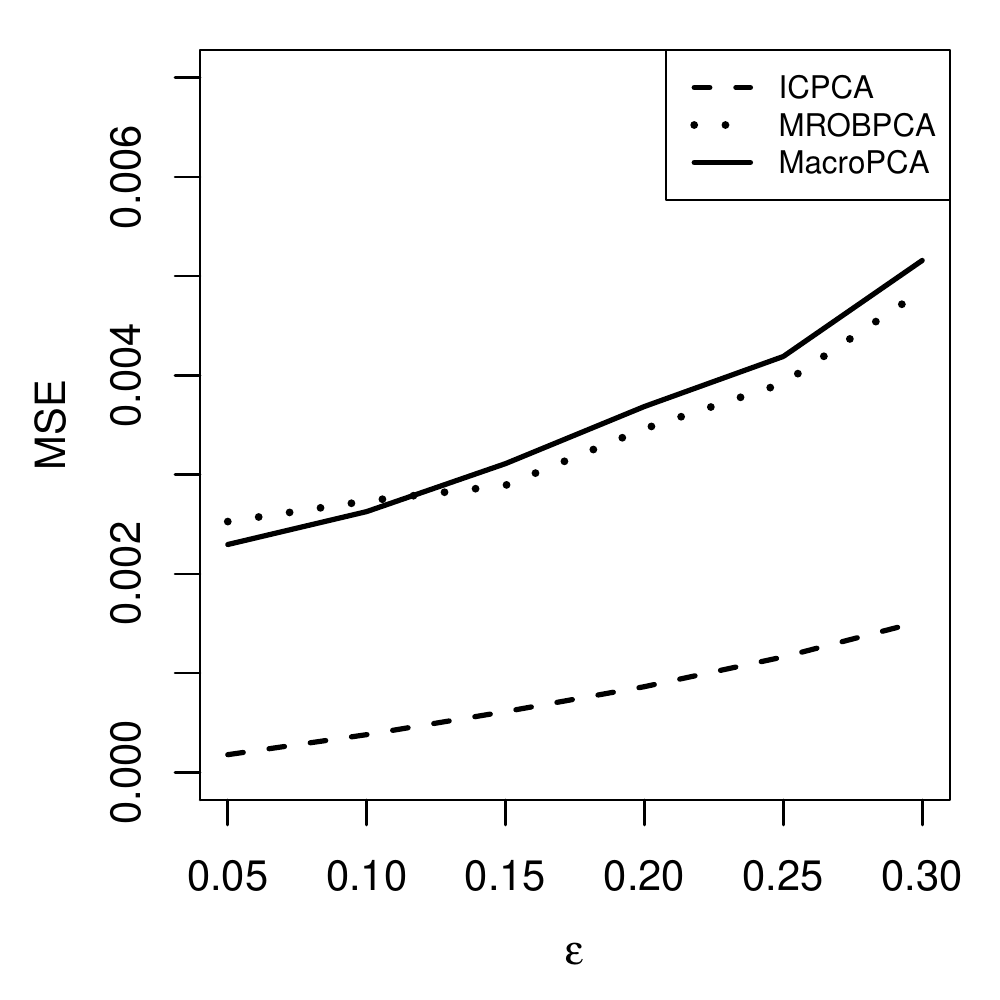} & 
 \includegraphics[width=0.45\textwidth]
	 {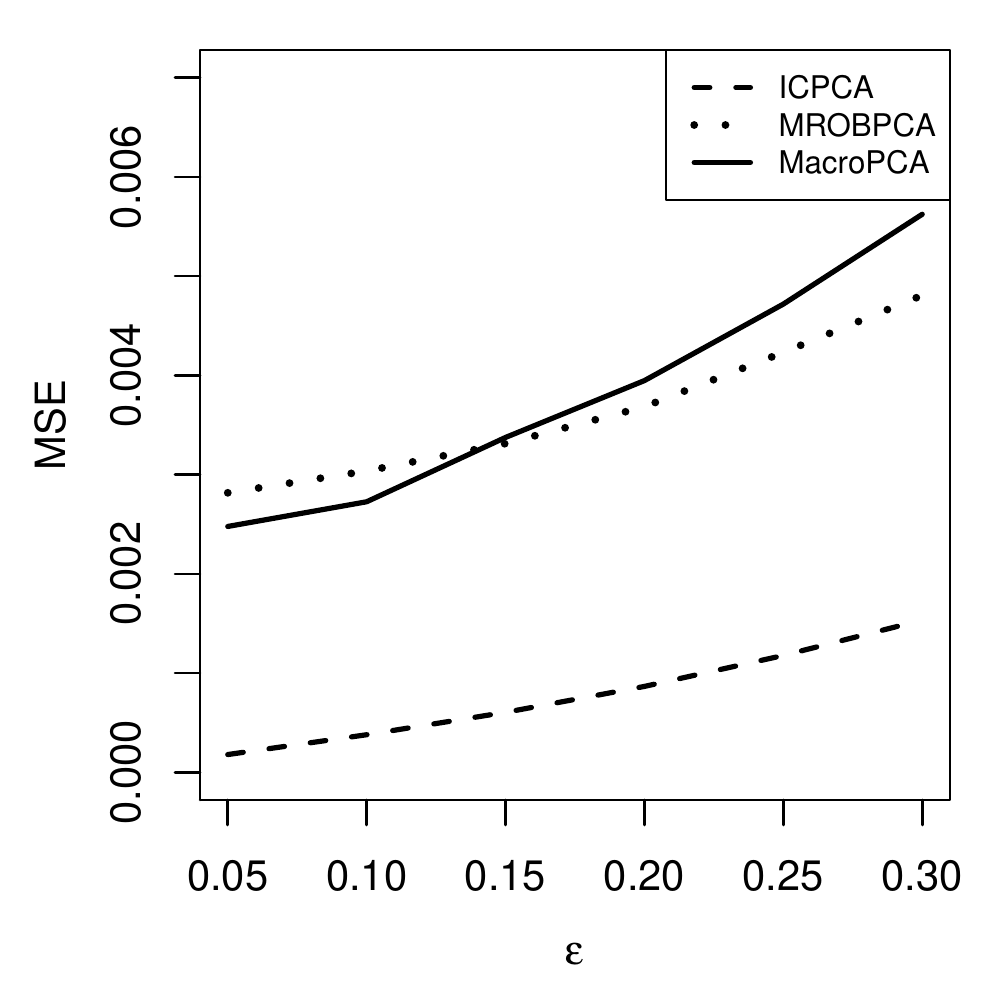}
\vspace{-0.3cm}	
\end{tabular}
\caption{Average MSE as a function of the fraction 
  $\eps$ of missing values.	The data were generated
	using A09 (left) and ALYZ (right).}
	\label{fig:MCAR}
\end{figure}

Figure \ref{fig:MCAR} shows the performance of 
ICPCA, MROBPCA and MacroPCA when some data becomes missing. 
As CPCA and ROBPCA cannot deal with NA's, they are 
not included in this comparison. 
Since there are no outliers the classical ICPCA performs
best, followed by MROBPCA and MacroPCA which perform
similarly to each other, and only slightly worse than
ICPCA considering the scale of the vertical axis which is 
much smaller than in the other three simulation settings. 
    
\begin{figure}[!ht]
\centering
\vskip2mm
\begin{tabular}{cc}
	\hskip7mm A09, missing values \& cellwise 
	& \hskip3mm ALYZ, missing values \& cellwise \\
		\includegraphics[width=0.45\textwidth]
	{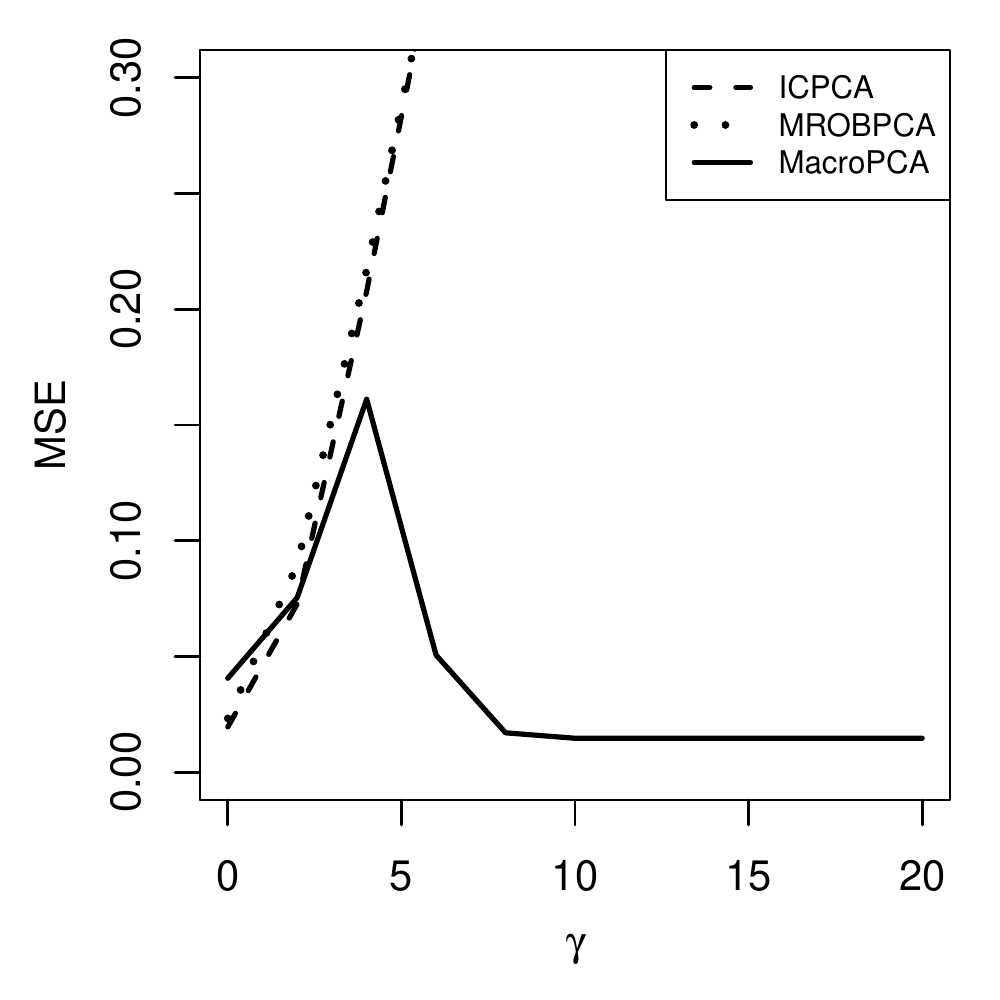} & 
		\includegraphics[width=0.45\textwidth]
	{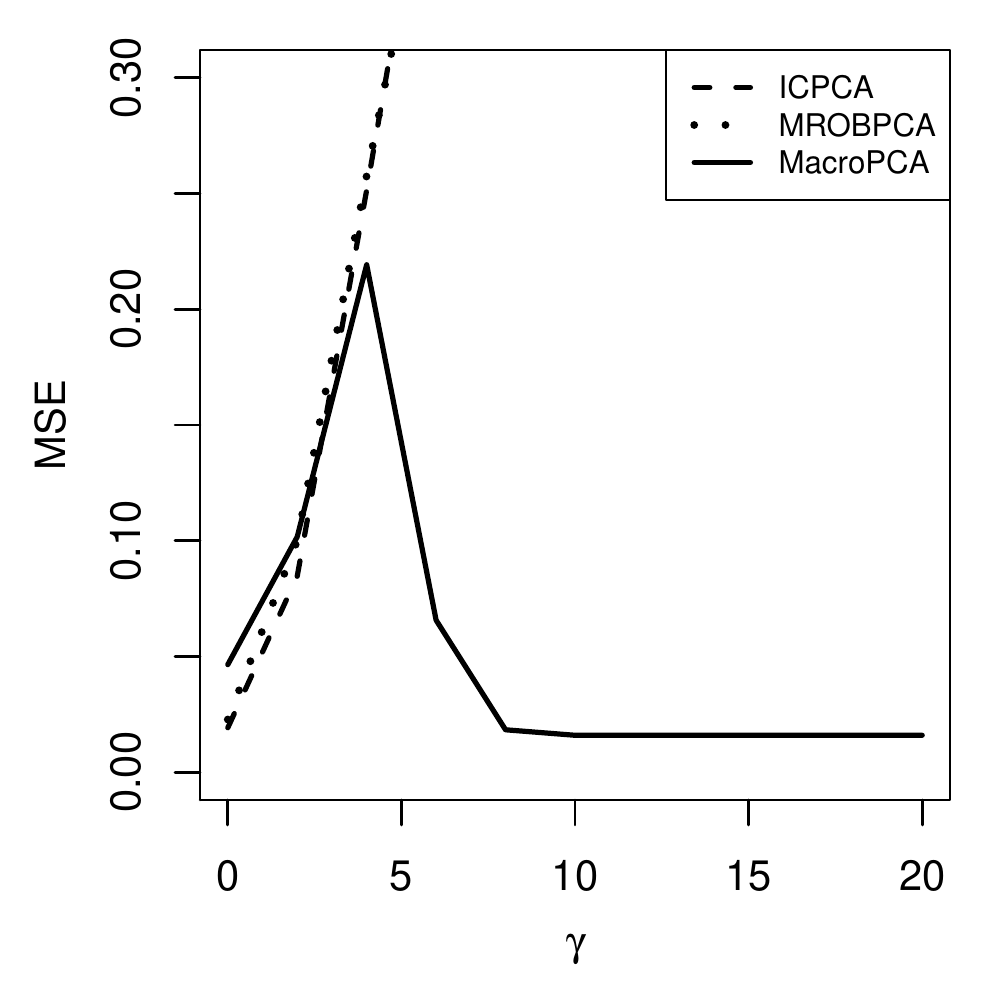}
\vspace{-0.2cm}
\end{tabular}
\caption{Average MSE for data with 20\% of missing
    values and 20\% of cellwise outliers, as a 
		function of $\gamma$ which determines the 
		distance of the cellwise outliers.} 
\label{fig:ICM}
\end{figure}

Now we set 20\% of the data cells to missing and 
add 20\% of cellwise contamination given by $\gamma$.
Figure~\ref{fig:ICM} shows the performance of ICPCA, 
MROBPCA and MacroPCA in this situation. 
The MSE of both ICPCA and MROBPCA grows very fast with
$\gamma$ which indicates that these methods are not at 
all robust to cellwise outliers. 
Note that $d=200$ so on average 
$1-(1-0.2)^{200}\approx 100 \%$ of the rows are 
contaminated, whereas no purely rowwise method 
can handle more than 50\%.
MacroPCA is the only method that can withstand cellwise
outliers here. When $\gamma$ is smaller than 5 the MSE
goes up, but this is not surprising as in that case the 
values in the contaminated cells are still close to the 
clean ones. As soon as the contamination is sufficiently 
far away, the MSE drops to a very low value. 

\begin{figure}[!ht]
\centering
\vskip2mm
\begin{tabular}{cc}
	\hskip8mm A09, missing values \& rowwise & 
	\hskip3mm ALYZ, missing values \& rowwise \\
		\includegraphics[width=0.45\textwidth]
	{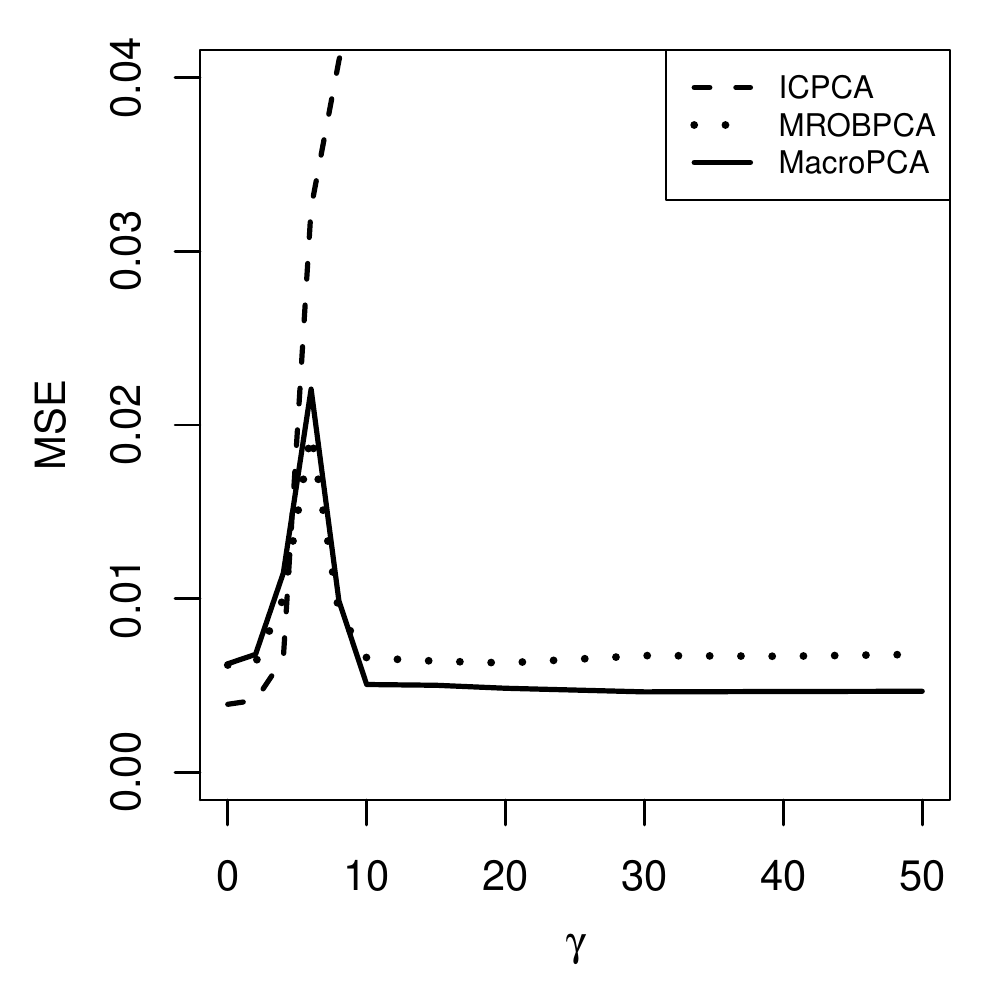} & 
		\includegraphics[width=0.45\textwidth]
	{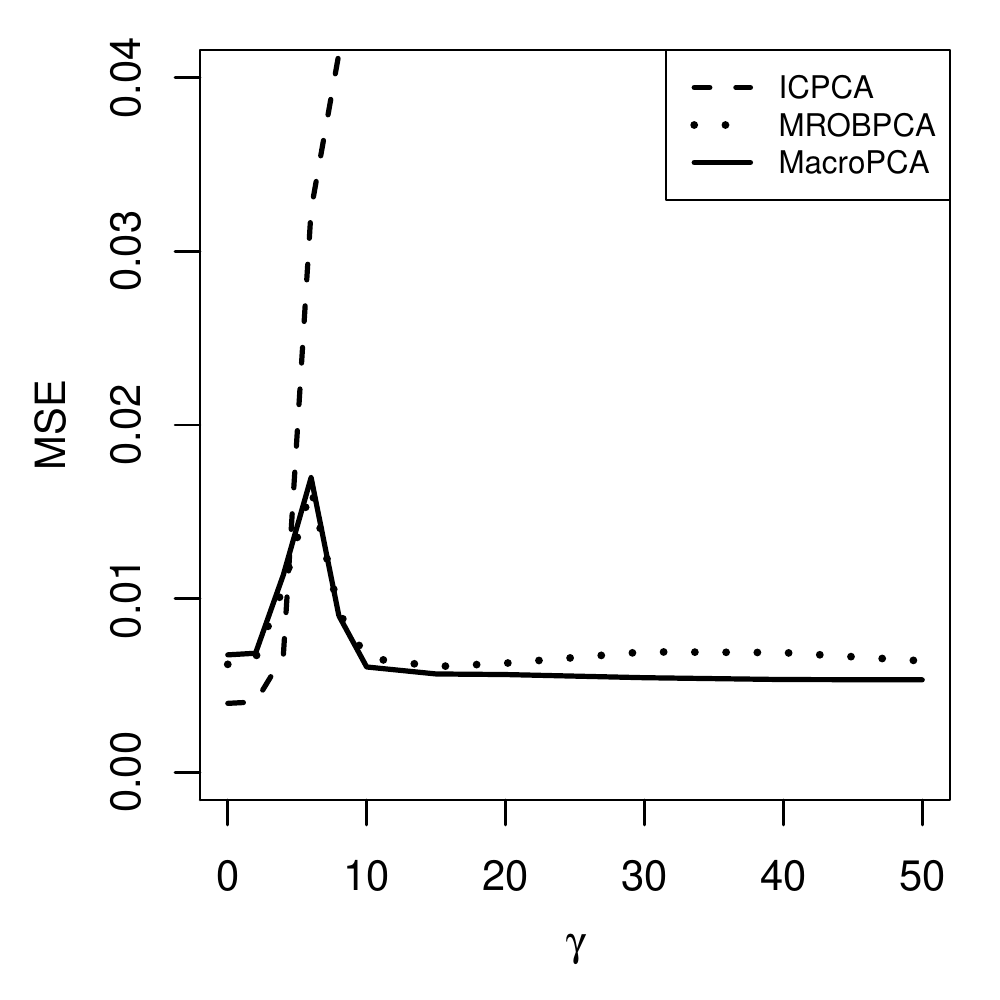}
\vspace{-0.2cm}
\end{tabular}
\caption{Average MSE for data with 20\% of missing
    values and 20\% of rowwise outliers, as a 
		function of $\gamma$ which determines the 
		distance of the rowwise	outliers.} 
\label{fig:THCM}
\end{figure}

Figure~\ref{fig:THCM} presents the results of ICPCA, 
MROBPCA and MacroPCA when there are 20\% of missing
values combined with 20\% of rowwise contamination. 
As expected, the ICPCA algorithm breaks down while
MROBPCA and MacroPCA provide very good results. 
MROBPCA and MacroPCA are affected the most (but 
not much) by nearby outliers, and very little by far 
contamination.    

\begin{figure}[!ht]
\centering
\vskip2mm
\begin{tabular}{cc}
\hskip 1mm A09, missing \& cellwise \& rowwise & 
\hskip 1mm ALYZ, missing \& cellwise \& rowwise \\
	\includegraphics[width=0.45\textwidth]
	{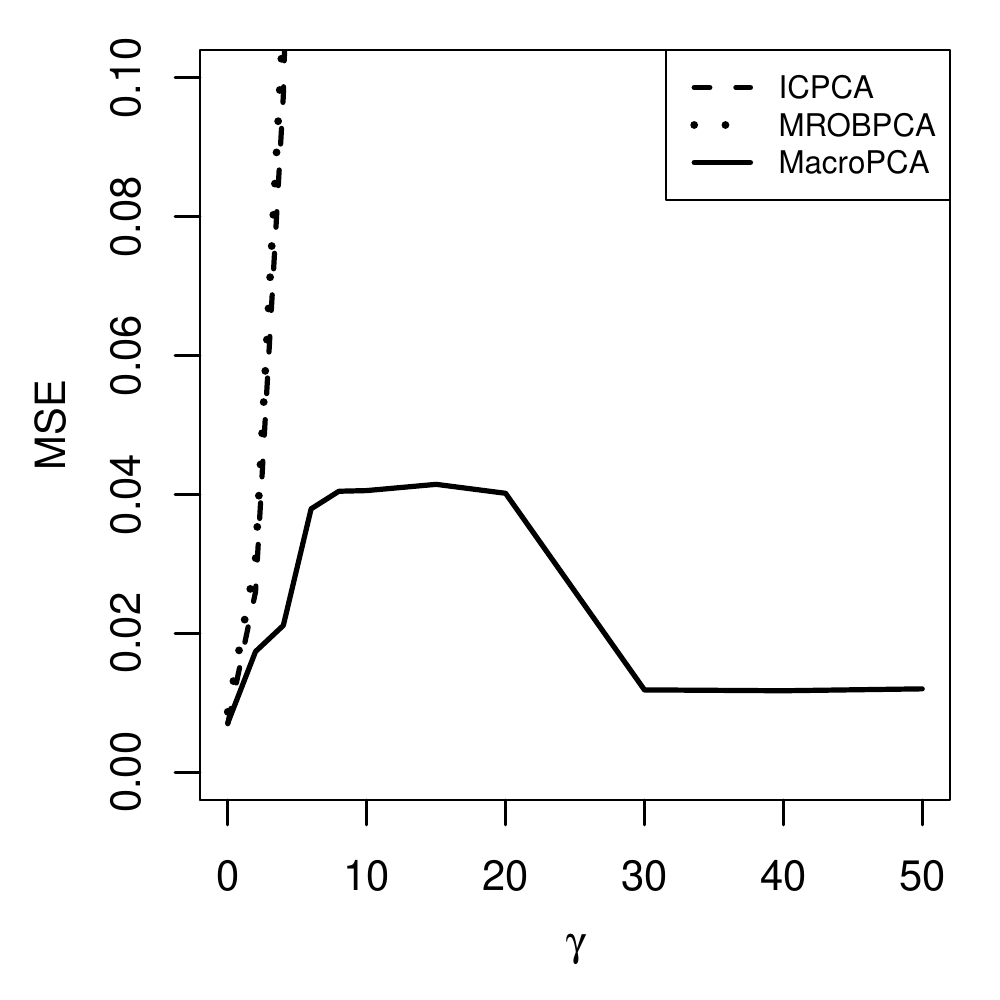} & 
	\includegraphics[width=0.45\textwidth]
	{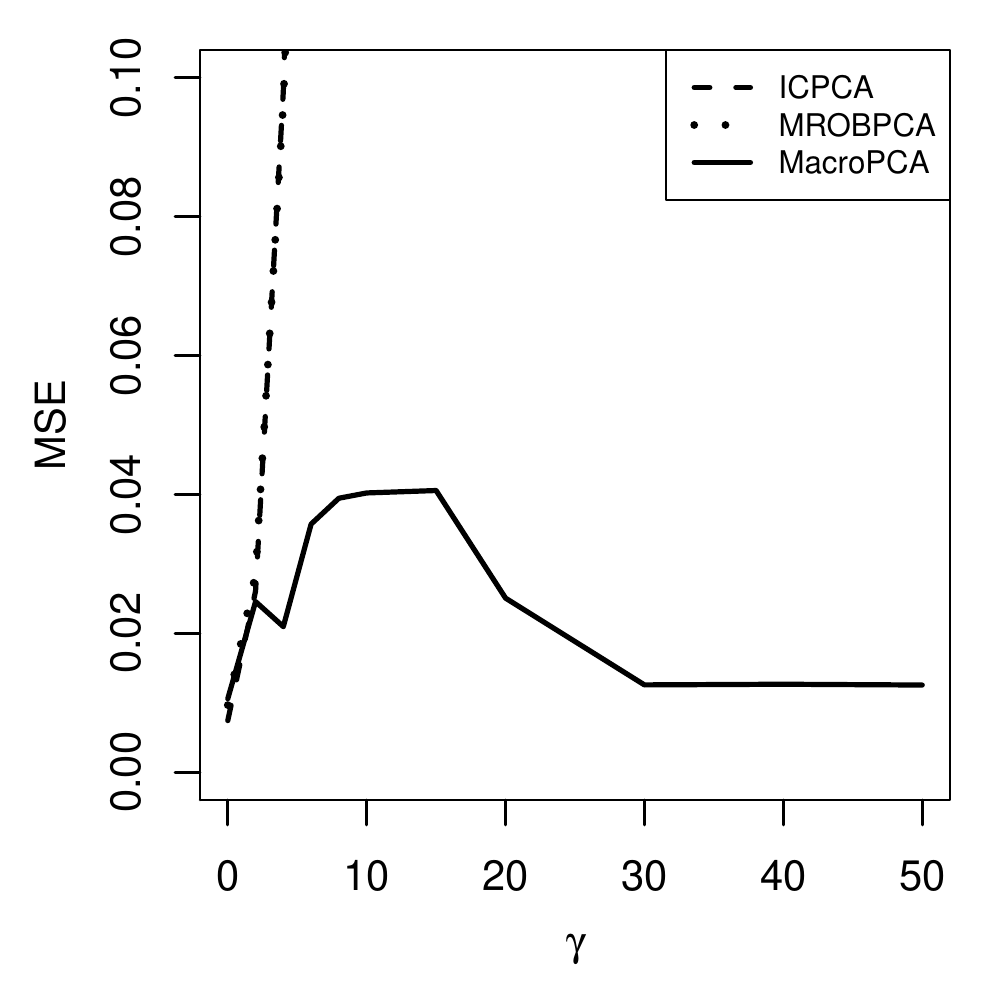}
\vspace{-0.2cm}
\end{tabular}
\caption{Average MSE for data with 20\% of missing
    values, 10\% of cellwise outliers and 10\% of 
		rowwise outliers, as a function of $\gamma$ 
		which determines the distance of both the 
		cellwise and the rowwise outliers.} 
\label{fig:BOTH}
\end{figure}

Finally, Figure~\ref{fig:BOTH} presents the results 
in the situation of 20\% of missing values combined 
with 10\% of cellwise and 10\% of rowwise 
contamination. 
In this scenario the ICPCA and MROBPCA algorithms 
break down whereas MacroPCA still provides reasonable
results.
 
In this section the missing values were generated
in a rather simple way. In Section \ref{A:MAR}
they are generated in a more challenging way but
still MAR, with qualitatively similar results.

\section{Real data examples}
\label{sec:Examples}

\subsection{Glass data}
\label{sec:glass}

\begin{figure}[htb]
\centering
\includegraphics[width=1.0\textwidth]
	{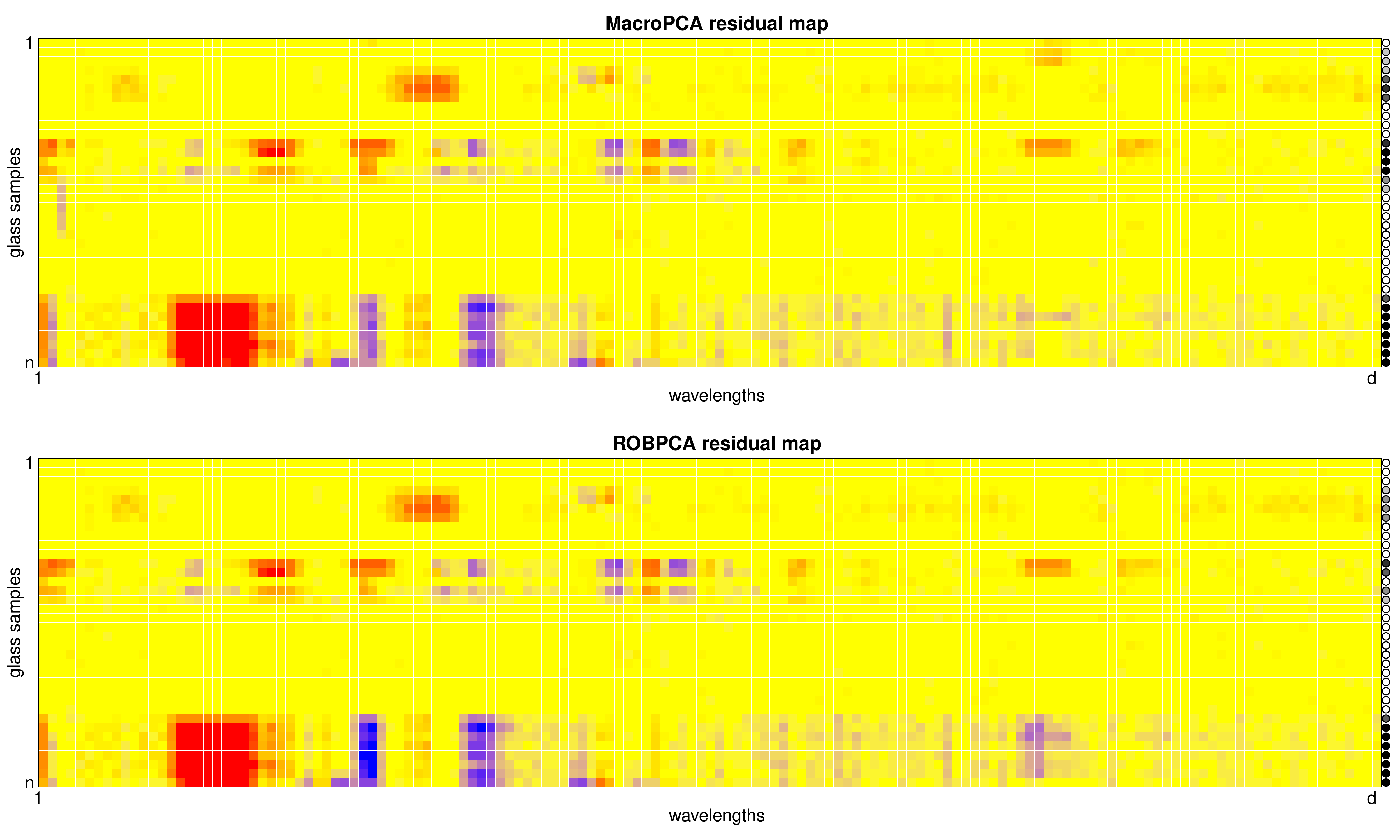}
\vspace{-1.0cm}
\caption{Residual maps of the glass dataset when 
  fitting the PCA model by MacroPCA (top) and 
	ROBPCA (bottom).}
\label{fig:Glass1}
\end{figure}

The {\it glass} dataset \citep{Lemberge:PLS} 
contains spectra with $d=750$ wavelengths of $n=180$ 
archeological glass samples. 
It is available in the R package
{\it cellWise} \citep{Raymaekers:cellWise}. 
The MacroPCA method selects 4 principal components
and yields a $180 \times 750$ matrix of 
standardized residuals. There is not enough 
resolution on a page to show so many individual 
cells in a residual map.
Therefore we created a map (the top panel of
Figure \ref{fig:Glass1}) which combines the residuals 
into blocks of $5 \times 5$ cells. 
The color of each block now depends on the most 
frequent type of outlying cell in it, the resulting
color being an average.
For example, an orange block indicates that quite a
few cells in the block were red and most of the 
others were yellow.
The more red cells in the block, the darker red the
block will be.
We see that MacroPCA has flagged a lot of cells,
that happen to be concentrated in a minority of the
rows where they show patterns.
In fact, the colors indicate that some of the glass 
samples (between 22 and 30) have a higher 
concentration of phosphor, whereas rows 57--63 and 
74--76 had an unusually high concentration of calcium.
The bottom part of the residual map looks very 
different,
due to the fact that the measuring instrument was 
cleaned before recording the last 38 spectra.
One could say that those outlying rows belong to a 
different population.

Since the dataset has no NA's and we found that
fewer than half of the rows are outlying, it can 
also be analyzed by the original ROBPCA method as 
was done by \cite{Hubert:ROBPCA}, also for $k = 4$.
This detects the same rowwise outliers.
In principle ROBPCA is a purely rowwise method that 
does not flag cells.
Even though ROBPCA does not produce a residual map, 
we can construct one analogously to that of MacroPCA.
First we construct the residual matrix of ROBPCA,
the rows of which are given by $\bx_i - \bhx_i$
where $\bhx_i$ is the projection of $\bx$ on the
ROBPCA subspace.
We then standardize the residuals in each
column by dividing them by a 
robust 1-step scale M-estimate.
This yields the bottom panel of 
Figure \ref{fig:Glass1}.
We see that the two residual maps look quite
similar.

This example illustrates that purely rowwise robust
methods can be useful to detect cellwise outliers 
when these cells occur in fewer than 50\% of the rows. 
But if the cellwise outliers contaminate more rows,
this approach is insufficient.

\subsection{DPOSS data}
\label{sec:DPOSS}

In our last example we analyze data from the 
Digitized Palomar Sky Survey (DPOSS) described by
\cite{Odewahn:Sky}.
This is a huge database of celestial objects,
from which we have drawn 20,000 stars at random.
Each star has been observed in the color bands
J, F, and N. 
Each band has 7 variables.
Three of them measure light intensity:
for the J band they are MAperJ, MTotJ and
MCoreJ where the last letter indicates the band. 
The variable AreaJ is the size of the star
based on its number of pixels.
The remaining variables IR2J, csfJ and EllipJ
combine size and shape.
(There were two more variables in the original
data, but these measured the background rather 
than the star itself.)
There are substantial correlations between these
21 variables.

\begin{figure}[!ht]
\centering
\vspace{0.3cm}
\begin{tabular}{cc}
\includegraphics[width=0.485\textwidth]
	{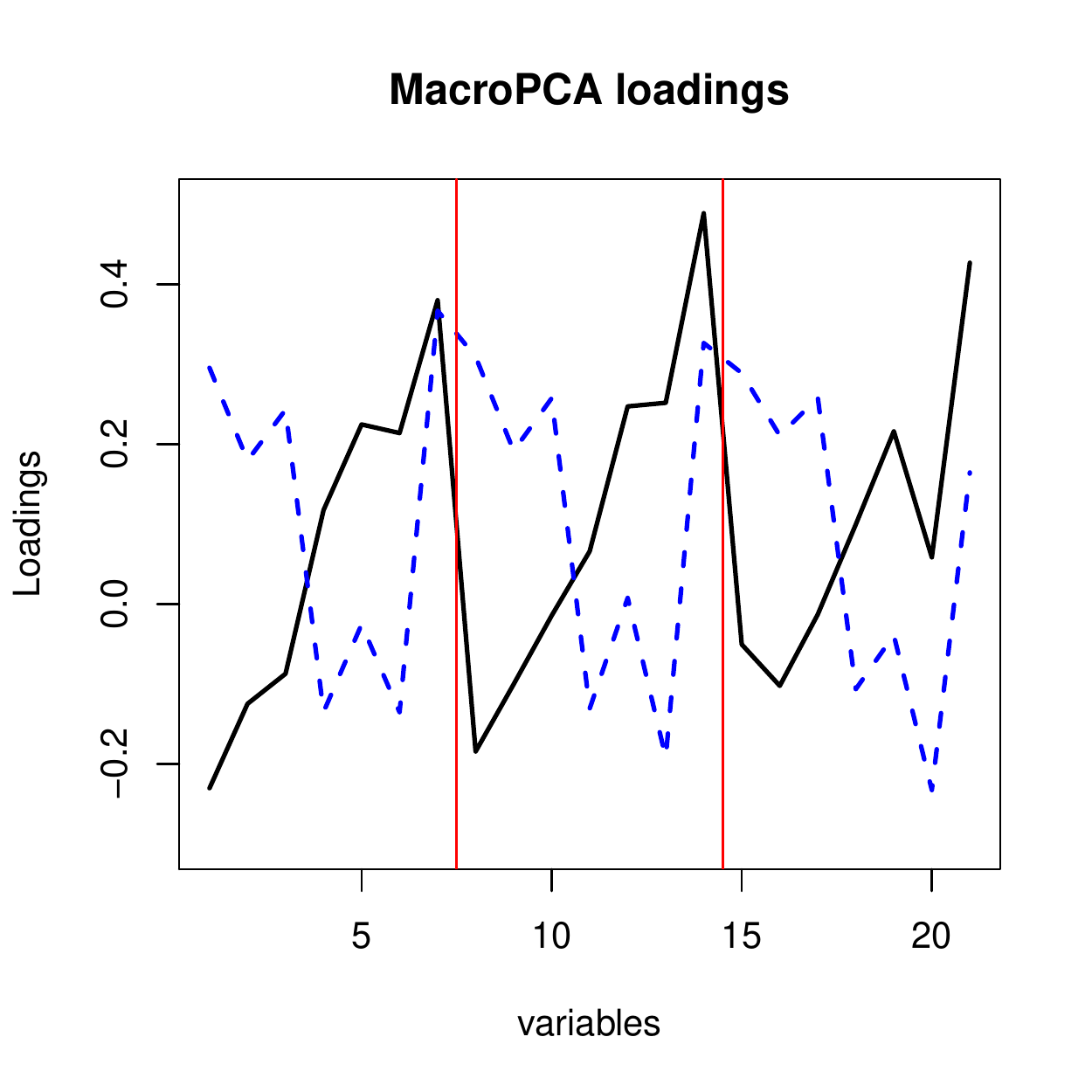} & 
\includegraphics[width=0.485\textwidth]
	{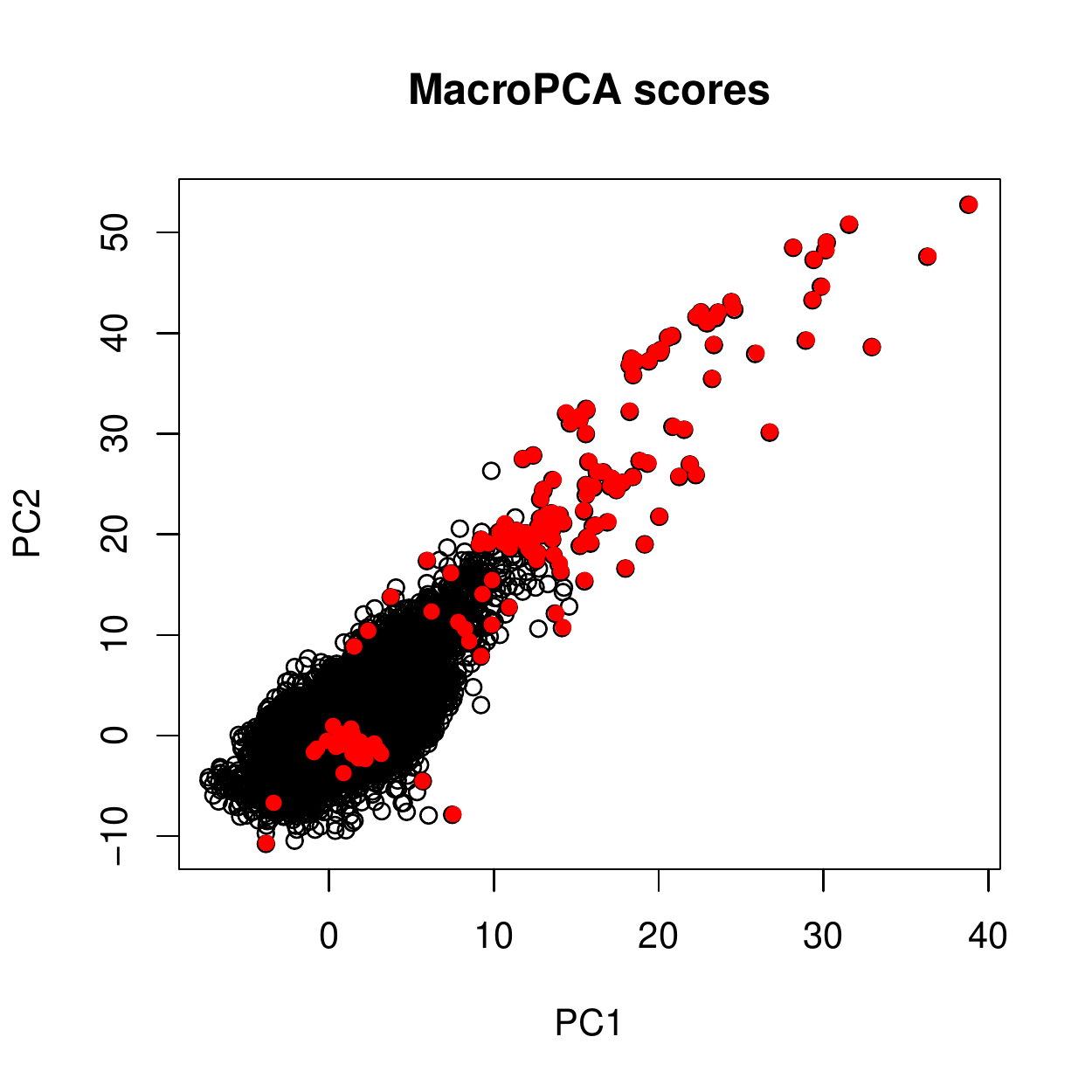}\\
\end{tabular}
\vspace{-0.8cm}
\caption{DPOSS stars data: (left) loadings of the 
  first (black full line) and the second (blue 
	dashed line) component of MacroPCA, with vertical 
	lines separating the three color bands; (right)
	plot of
	the first two scores, with filled red circles
	for stars with high orthogonal distance 
	$\protect\cod$ and
	open black circles for the others.}
\label{fig:DPOSSloadings}
\end{figure}

In this dataset 84.6\% of the rows contain NA's
(in all there are 50.2\% missing entries.) 
Often an entire color band is missing, and
sometimes two. 
We applied MacroPCA to these data, choosing $k=4$
components according to the scree plot.
The left panel of Figure \ref{fig:DPOSSloadings} 
shows the loadings of the first and second 
component.
It appears that the first component captures the
overall negative correlation between two groups
of variables: those measuring light intensity
(the first 3 variables in each band) and the
others (variables 4 to 7 in each band). 
The right panel is the corresponding scores plot, 
in which the 150 stars with the
highest orthogonal distance $\cod$ are shown in red.
Most of these stand out in the space of PC1 and 
PC2 (bad leverage points), whereas some only
have a high $\cod$ (orthogonal outliers).

\begin{figure}[!ht]
\centering
\includegraphics[width=0.7\textwidth]
	{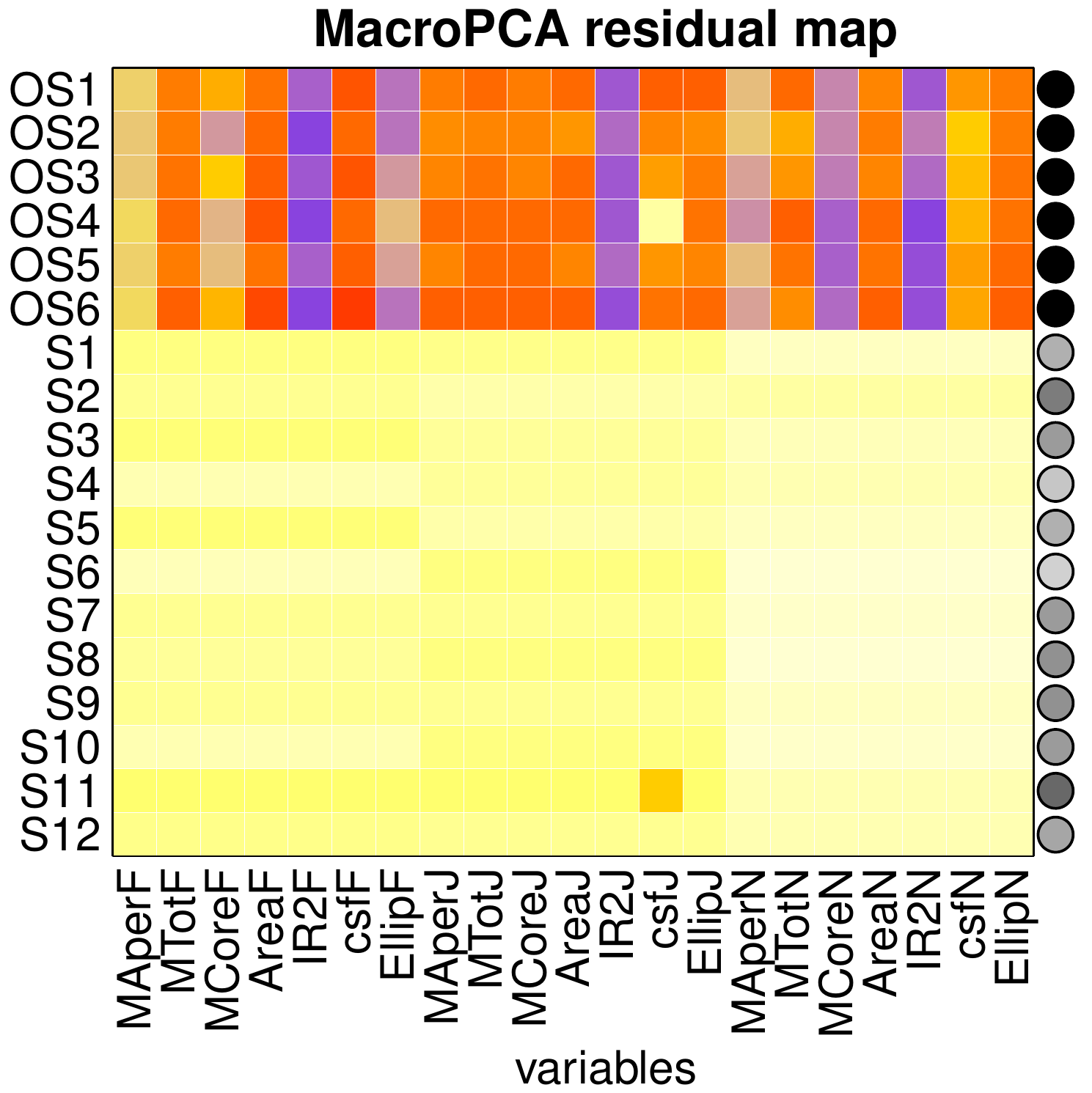}
\vspace{-0.5cm}
\caption{MacroPCA residual map of stars in the 
  DPOSS data, with 25 stars per row block.
  The six row blocks at the top correspond
	to the stars with highest 
	$\protect\cod$.}
\label{fig:DPOSScellmap}
\end{figure}

Figure \ref{fig:DPOSScellmap} shows the residual
map of MacroPCA, in which each row 
block combines 25 stars. 
The six rows at the top correspond to the  
150 stars with highest $\cod$.
We note that the outliers tend to be more
luminous (MTot) than expected
and have a larger Area, which suggests
giant stars.
The analogous residual map of ICPCA (not shown)
did not reveal much.
Note that the non-outlying rows in the bottom 
part of the residual map are yellow, and the
missing color bands show up as blocks in 
lighter yellow (a combination of yellow and 
white cells).

\section{Conclusions}
\label{sec:Conclusions}
The MacroPCA method is able to handle missing 
values, cellwise outliers, and rowwise outliers.
This makes it well-suited for the analysis of 
possibly messy real data.
Simulation showed that its performance is similar 
to a classical method in the case of outlier-free 
data with missing values, and to an existing 
robust method when the data only has rowwise 
outliers. 
The algorithm is fast enough to deal with many 
variables, and we intend to speed it up
by recoding it in C.

MacroPCA can analyze new data as they come in,
only making use of its existing output obtained
from the initial dataset.
It imputes missing values in the new data, 
flags and imputes outlying cells, and flags 
outlying rows.
This computation is fast, so it can be used to 
screen new data in quality control or even
online process control. (One can update the
initial fit offline from time to time.)
The advantage of MacroPCA is that it not only 
tells us when the process goes out of control, 
but also which variables are responsible.

Potential extensions of MacroPCA include methods
of PCA regression and partial least squares 
able to deal with rowwise and 
cellwise outliers and missing values.

\section{Software Availability}
\label{sec:software}
The R code of MacroPCA, as well as the data
sets and an example script, are available at
{\it https://wis.kuleuven.be/stat/robust/software}.
It will be incorporated in the R package
{\it cellWise} on CRAN.

\appendix
\numberwithin{equation}{section} 
\section{Appendix} \label{sec:A}
\renewcommand{\theequation}
   {\thesection.\arabic{equation}}

\subsection{Example illustrating step 5 
            of MacroPCA}
\label{A:toy}

Steps 1 to 4 of the MacroPCA algorithm construct
a robust $k$-dimensional subspace. 
The purpose of step 5 is to robustly estimate 
basis vectors (loadings) of the subspace. 
The columns of the loadings matrix $\bP_{d,k}^*$ 
from step 4 need not be robust, because some of 
the $n^*$ rows in $H^*$ might be outlying inside 
the subspace. 
Such points are called good leverage points,
where `good' refers to the fact that they do not 
harm the estimation of the subspace, but on the
other hand they can affect the estimated 
eigenvectors and eigenvalues.
We illustrate this by a small toy example.

A clean data set with $n=100$ and $d=3$ was 
generated according to a trivariate Gaussian 
distribution, with covariance matrix chosen in 
such a way that $k=2$ components explain 90\% of 
the total variance. 
Next, the data was contaminated by replacing
20\% of the points by good leverage points. 
This means that these points are only outlying 
inside the two-dimensional subspace, and not in 
the direction of its orthogonal complement.

\begin{figure}[!ht]
\centering
\vspace{0.5cm}
\includegraphics[width=0.55\textwidth] 
	{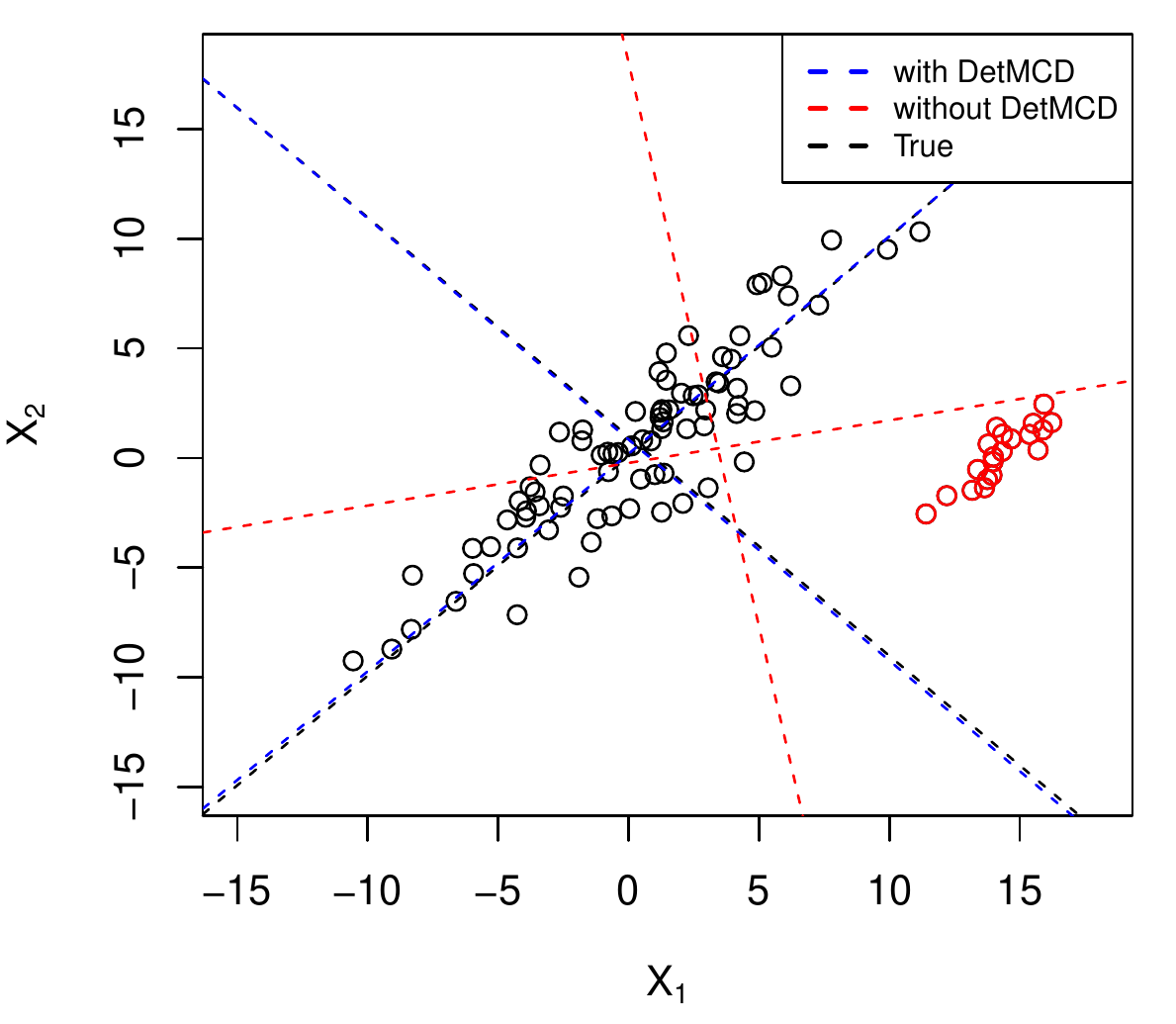}
\vspace{-0.5cm}
\caption{Scatter plot of the contaminated
data set projected on the subspace with $k=2$.
The blue points are from the clean data, and
the red ones are good leverage points. 
The dashed lines represent the eigenvectors.}
\label{fig:DetMCD}
\end{figure}

\begin{figure}[!ht]
\centering
\vskip2mm
\begin{tabular}{cc}
\hskip 1mm With step 5 & 
\hskip 1mm Without step 5\\
\includegraphics[width=0.47\textwidth]
 	{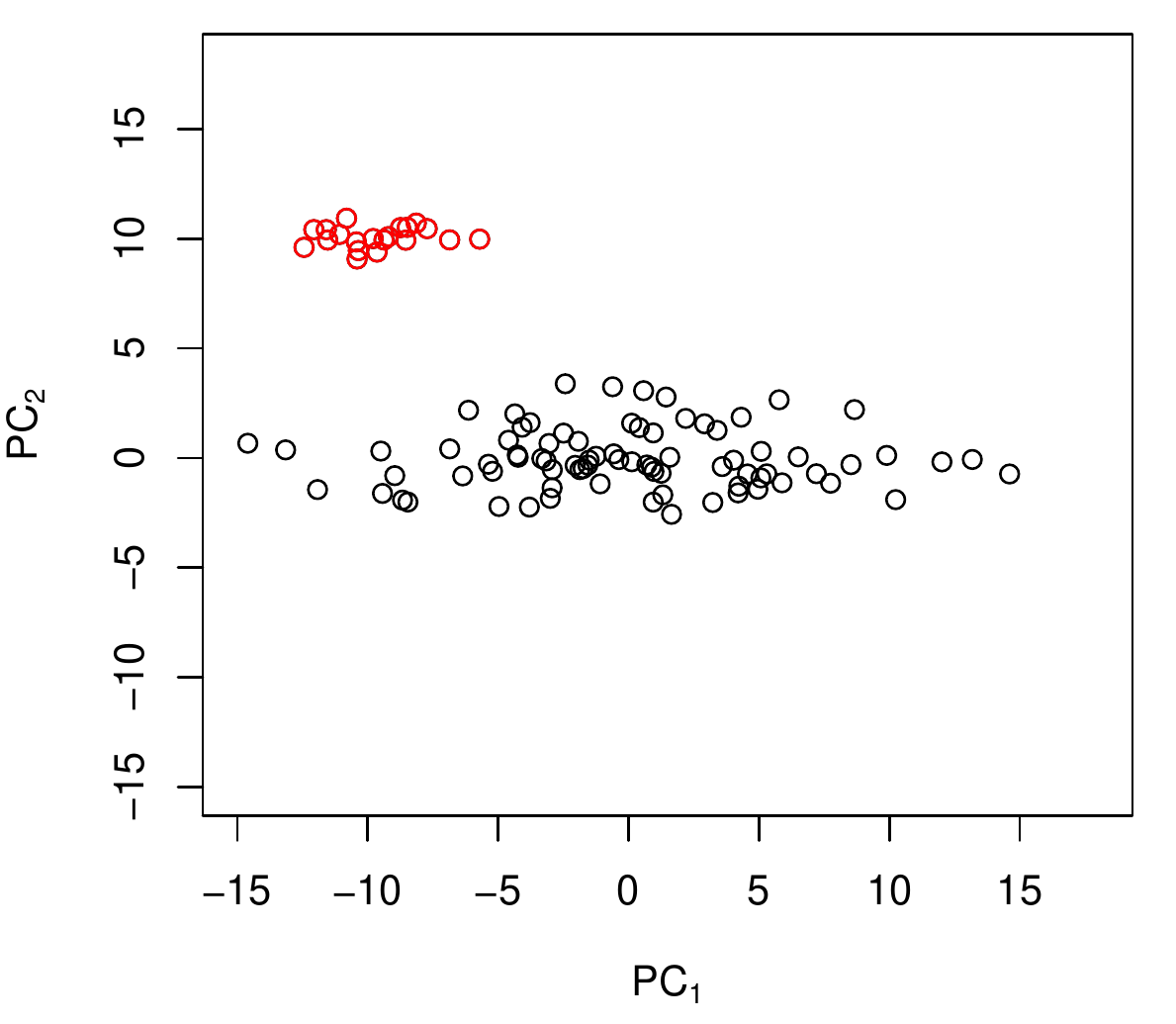} & 
\includegraphics[width=0.47\textwidth]
 	{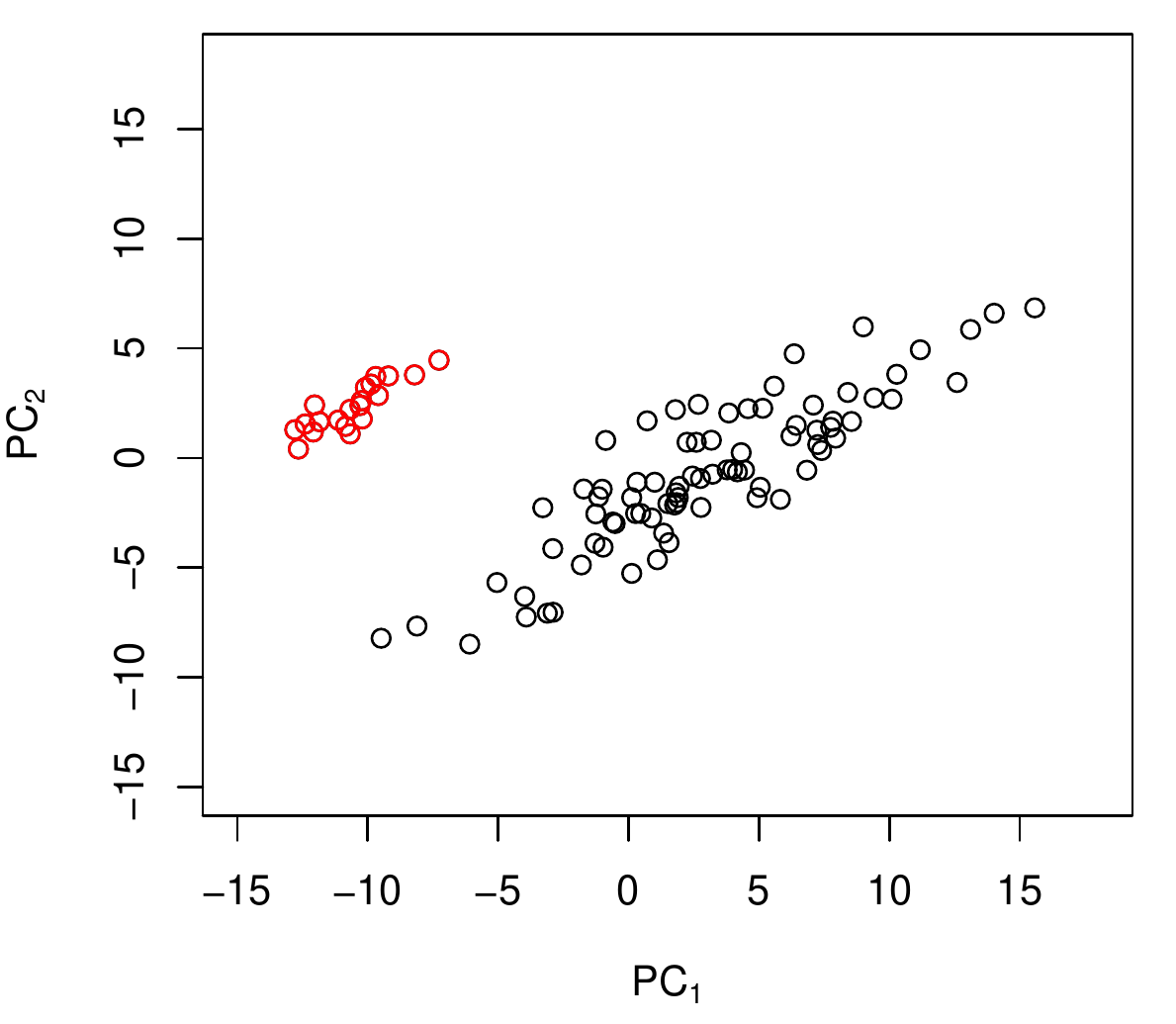}
\vspace{-0.4cm}
\end{tabular}
\caption{Scores plots with step 5 (left) 
  and without it (right).} 
\label{fig:scores_DetMCD}
\end{figure}

Figure \ref{fig:DetMCD} shows the scatter plot 
of the data points projected on this 
two-dimensional subspace. 
The black dashed lines represent the eigenvectors 
of the clean data, and the blue and red dashed 
lines are the estimated eigenvectors of MacroPCA 
on the contaminated data, with and without step 5.
The center and directions of the blue lines were
obtained by DetMCD in step 5, whereas the red
lines correspond to the center $\bm_d^*$ 
and the loading matrix $\bP_{d,k}^*$ obtained
in step 4.
The blue lines (with step 5) are almost on top of 
the black ones, whereas the red lines 
(without step 5) are very different.
Although the blue and the red fits span
the same subspace, we see that without step 5
the estimated loading vectors are far off.

This can also be seen in Figure 
\ref{fig:scores_DetMCD} which shows the scores 
on the first two components of MacroPCA with 
and without step 5. 
In the left panel we see that the scores
reflect the shape of the clean data,
whereas those in the right panel were 
affected by the leverage points.
This illustrates that the interpretation of the 
results can be distorted by good leverage 
points if step 5 is not taken.

\subsection{Computational complexity 
             of MacroPCA}
\label{A:complexity}

The computational complexity of MacroPCA
depends on how its steps are
implemented. The DDC in the first part has
an $O(nd\log(d))$ implementation 
\citep{Raymaekers:cellWise}.
The outlyingness \eqref{outlo} in 
step 1 requires $O(nd + n\log(n))$ time.
The classical PCA in steps 2, 3, and 4
needs $O(nd\min(n,d))$ if it is carried
out by singular value decomposition (SVD)
instead of the eigendecomposition of a 
covariance matrix.
The scores and predictions in steps 3, 4 and 6 
require $O(ndk)$ but since we assume that the 
number of components $k$ is at most 10 this 
becomes $O(nd)$. 
The DetMCD in step 5 combines 
initial estimators of total complexity 
$O(n\log(n)k^2)$ with C-steps that require
$O(n k^2)$ if performed by SVD. 
The robust scales of the residuals in step 6
take $O(n\log(n)d)$ time.
The overall complexity of MacroPCA thus becomes 
$O(nd(\min(n,d) +\log(n) + \log(d)))$ 
which is not much higher than the 
$O(nd\min(n,d))$ of classical PCA.

\begin{figure}[!ht]
\centering
\vskip2mm
\includegraphics[width=0.45\textwidth]
  {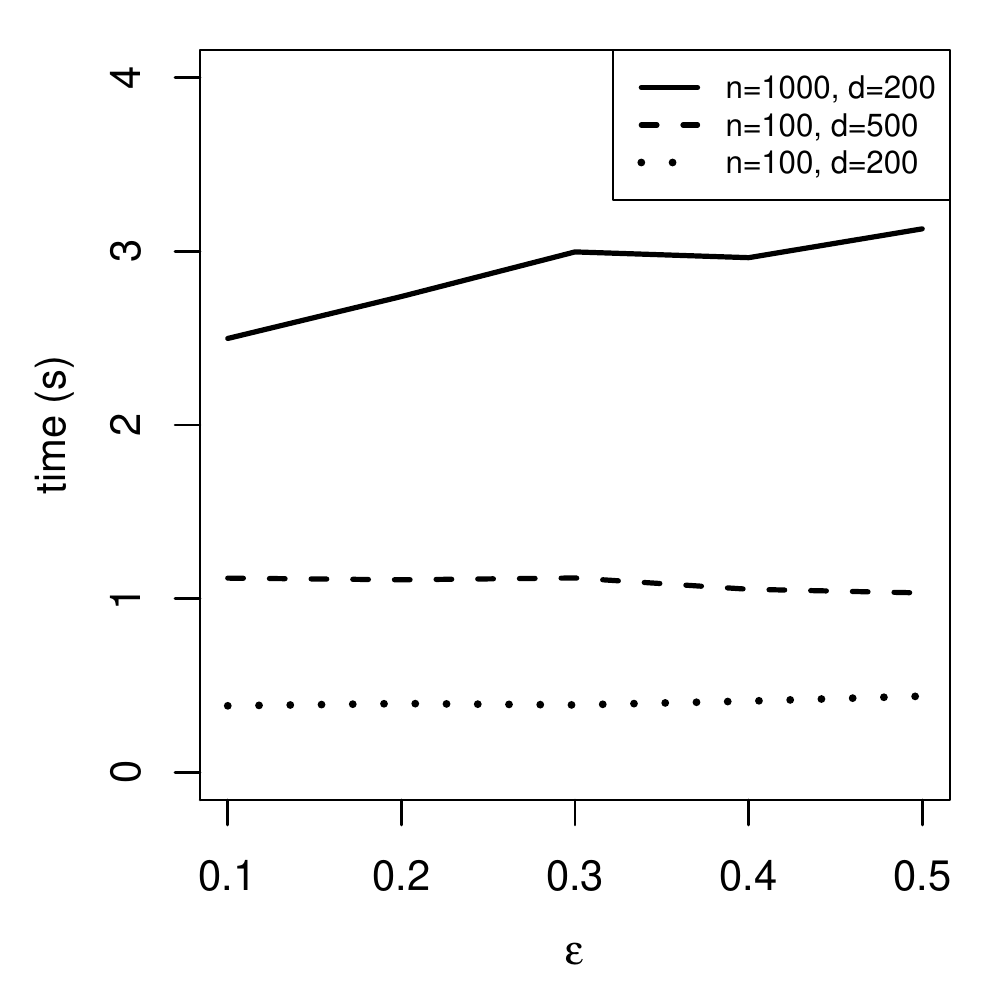} 
\vspace{-0.4cm}	
\caption{Computation times of MacroPCA in
  seconds, as a function of the fraction
	$\varepsilon$ of NA's in the data.}
	\label{fig:timesNAs}
\end{figure}

The computation times for a range of
values of $n$ and $d$ are shown in
Figure \ref{fig:times} in the paper.
The fraction of NA's in the data had no 
substantial effect on the computation 
time, as seen from Figure 
\ref{fig:timesNAs}.

\subsection{Generating NA's by a different MAR
            mechanism}
\label{A:MAR}

In this paper we have assumed that the missingness
mechanism is MAR (missing at random), 
a typical assumption underlying EM-based methods 
such as ICPCA and MROBPCA that are incorporated 
in MacroPCA.
MAR says that the missingness of a cell is
unrelated to the value the cell would have had,
but may be related to the values of other
cells in the same row; see, e.g., 
\cite{Schafer:missing}.

In the simulations of 
Section \ref{sec:Simulations} 
a random subset of $\varepsilon\%$ of the 
$n \times d$ cells was replaced by NA's.
This is a simple special case of MAR, in fact it
is MCAR (missing completely at random) which
assumes that the missingness of a cell does not
depend on its own value or that of any other
cell in its row.

Here we will look at a more challenging
mechanism that is still MAR but no longer MCAR.
Many MAR mechanisms are possible. We chose
the following one where the positions of the 
NA's clearly depend on the values of the other
cells. Given the uncontaminated matrix $\bX$ we 
first construct the matrix $\bU$ with cell values
\begin{equation}
\setlength{\jot}{5pt}
	\begin{aligned}
	u_{ij} & = |x_{id}|+|x_{i(j+1)}| & 
	 & \text{if} \ j = 1, \nonumber \\
	& = |x_{i(j-1)}|+|x_{i(j+1)}| & 
	& \text{if} \ 1<j<d, \nonumber \\ 
	& = |x_{i(j-1)}|+|x_{i1}| & 
	& \text{if} \ j = d. \nonumber
	\end{aligned}
\end{equation}

Next, the cells of $\bX$ corresponding to the 
$\varepsilon \%$ highest values of $\bU$ are 
set to missing. 
Other than this, the simulation setup is as 
described in Section \ref{sec:Simulations}.

\begin{figure}[!ht]
\centering
\vskip2mm
\begin{tabular}{cc}
\hskip5mm A09, fraction $\varepsilon$ of missing 
  values & \hskip0mm ALYZ, fraction $\varepsilon$ 
	of missing values \\
\includegraphics[width=0.45\textwidth]
   {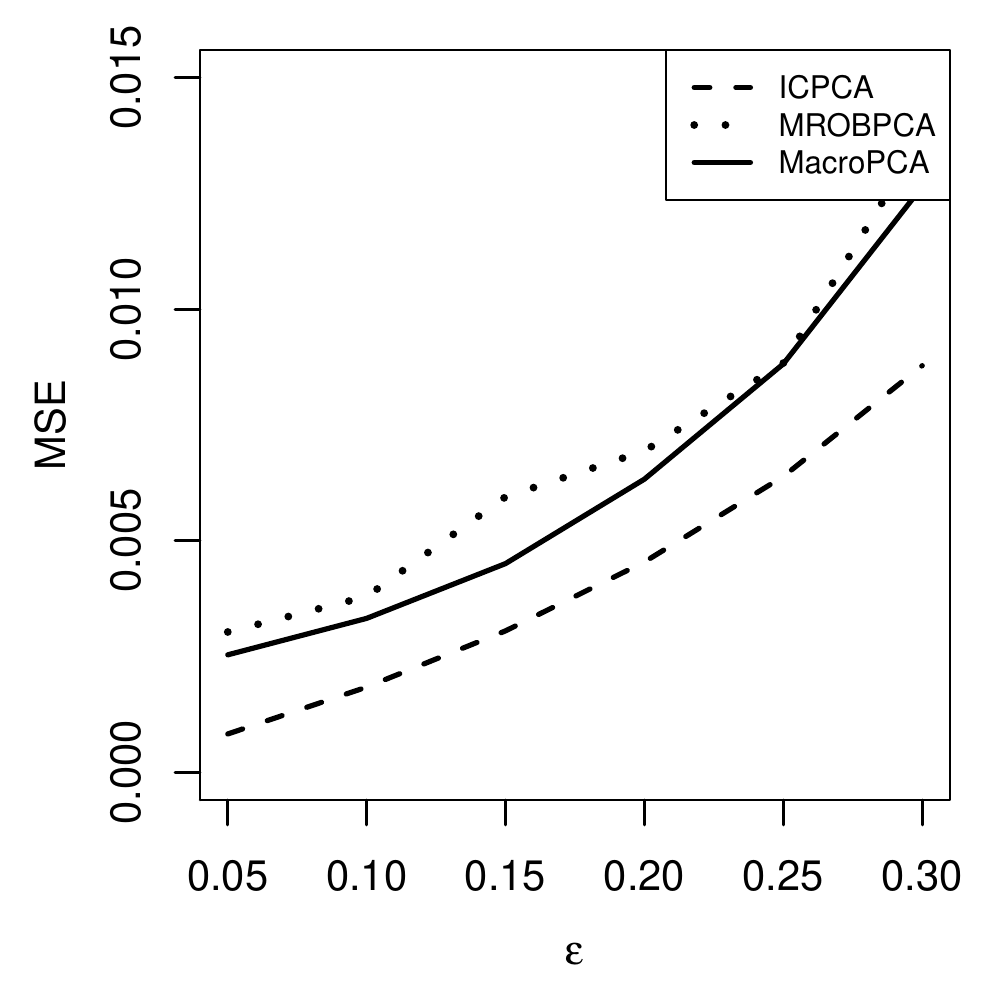} & 
\includegraphics[width=0.45\textwidth]
	 {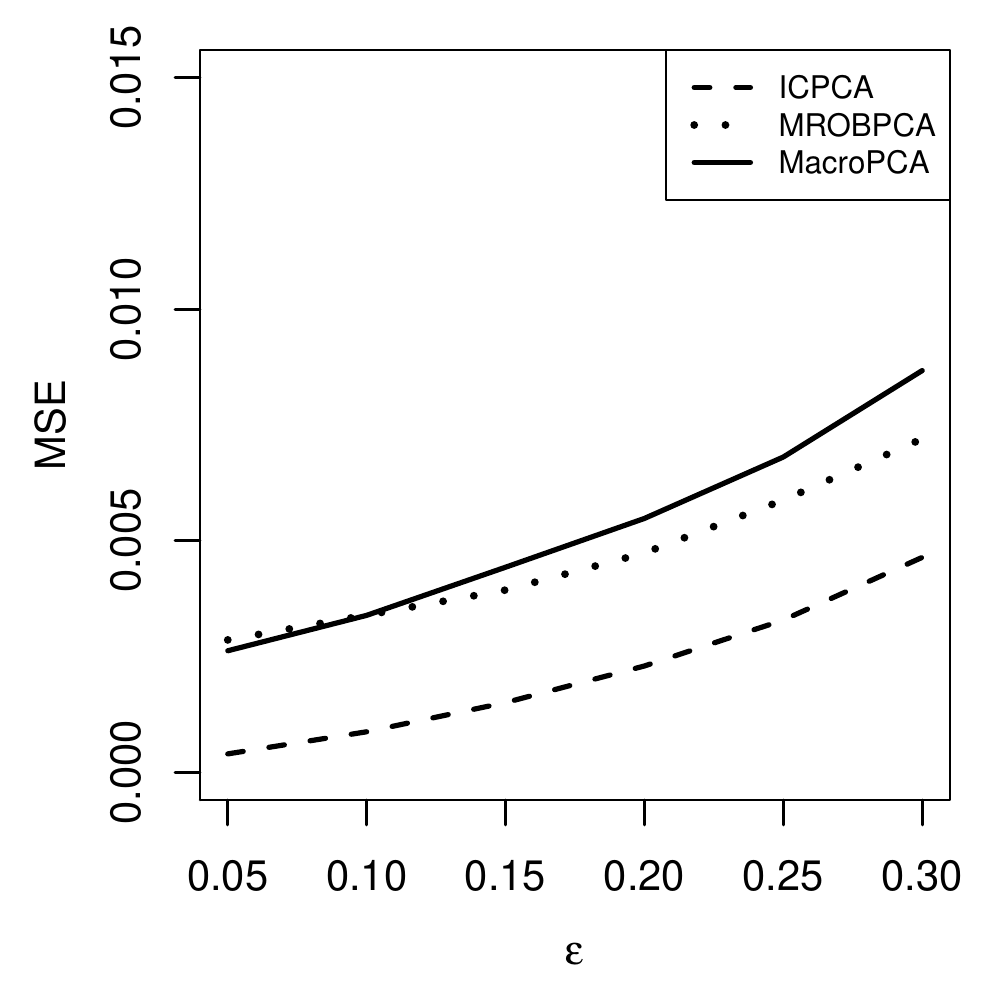}
\vspace{-0.3cm}	
\end{tabular}
\caption{Average MSE for data with MAR missingness 
  as a function of the fraction $\eps$ of missing 
	cells.}
\label{fig:MAR}
\end{figure}

The simulation results of ICPCA, MROBPCA and 
MacroPCA are shown in Figure \ref{fig:MAR}. 
Compared to the results with MCAR missing values 
in Figure \ref{fig:MCAR} we see that the shape 
of the plots is very similar, only the scale of
the MSE values is increased but it remains very 
low. 
    
\begin{figure}[!ht]
\centering
\vskip2mm
\begin{tabular}{cc}
	\hskip7mm A09, missing values \& cellwise 
	& \hskip3mm ALYZ, missing values \& cellwise \\
		\includegraphics[width=0.45\textwidth]
	{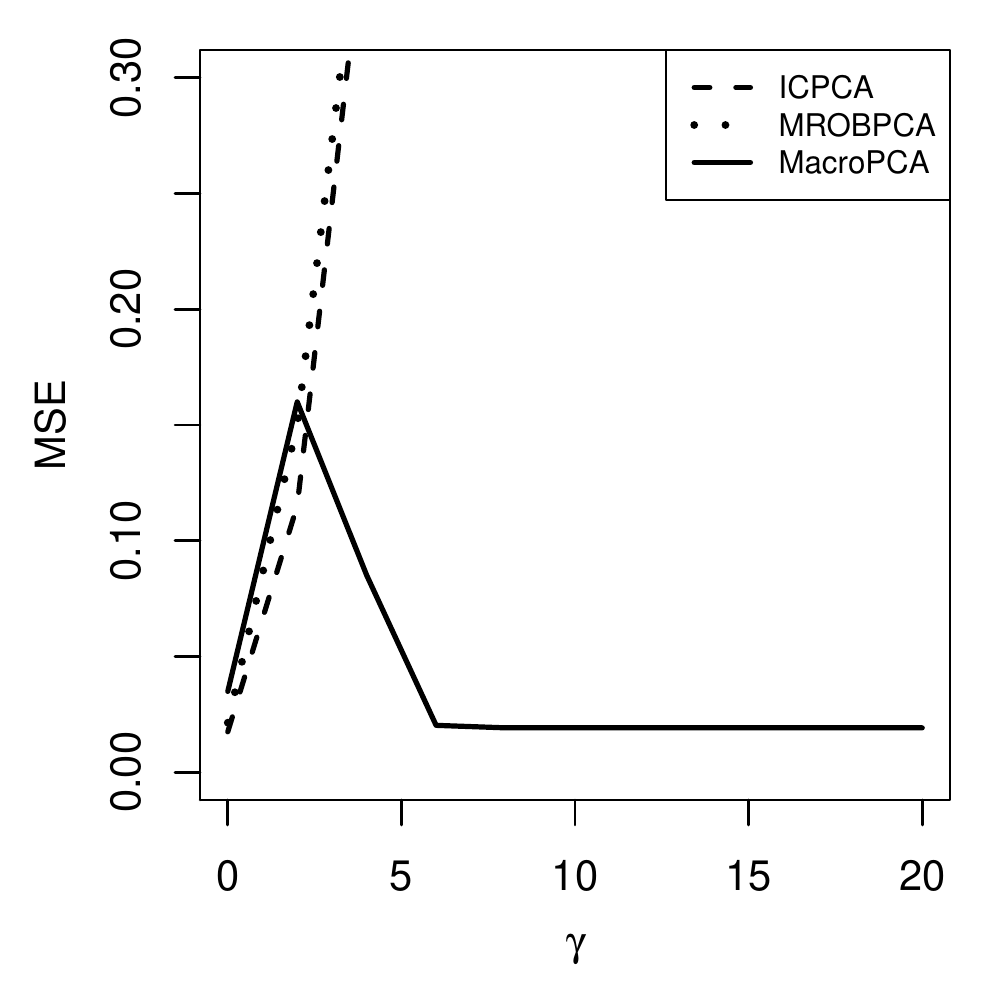} & 
		\includegraphics[width=0.45\textwidth]
	{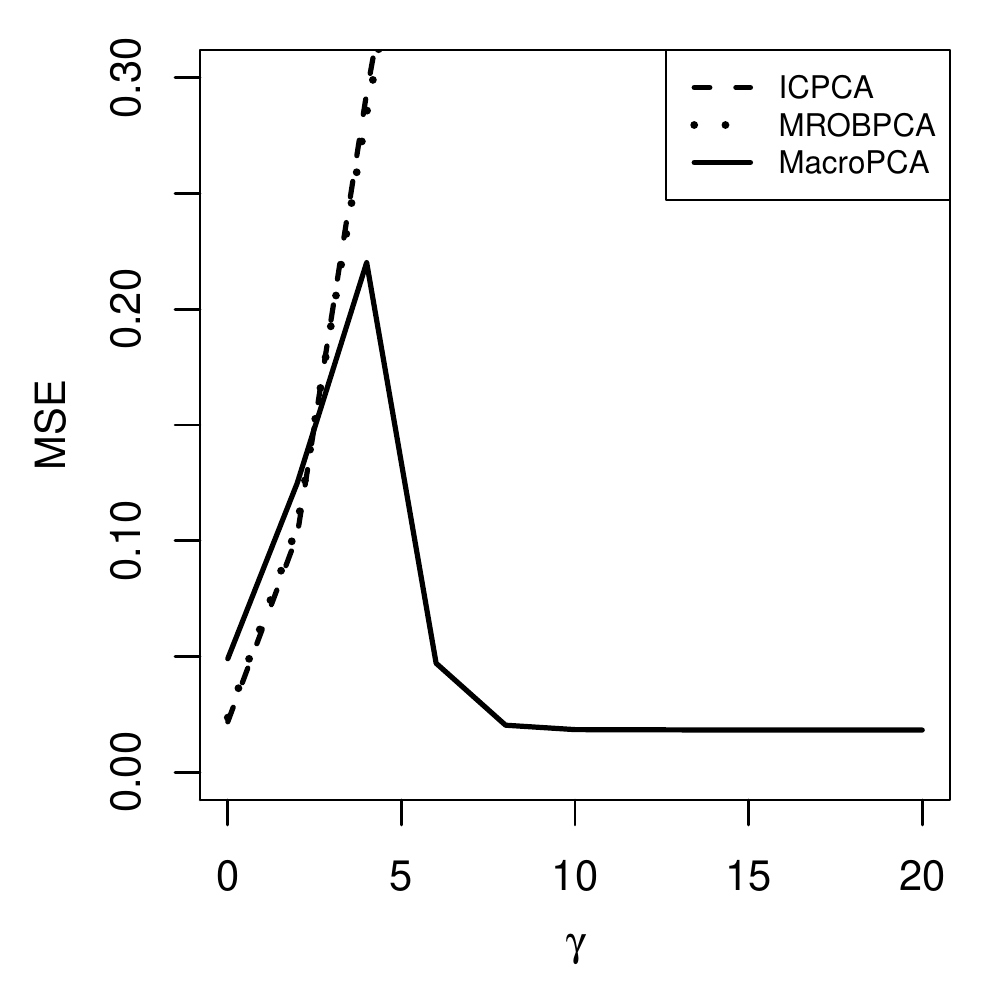}
\vspace{-0.2cm}
\end{tabular}
\caption{Average MSE for data with 20\% of MAR
    NA's and 20\% of cellwise outliers, as a 
		function of $\gamma$ which determines the 
		distance of the cellwise outliers.} 
\label{fig:ICMMAR}
\end{figure}

\begin{figure}[!ht]
\centering
\vskip2mm
\begin{tabular}{cc}
	\hskip8mm A09, missing values \& rowwise & 
	\hskip3mm ALYZ, missing values \& rowwise \\
		\includegraphics[width=0.45\textwidth]
	{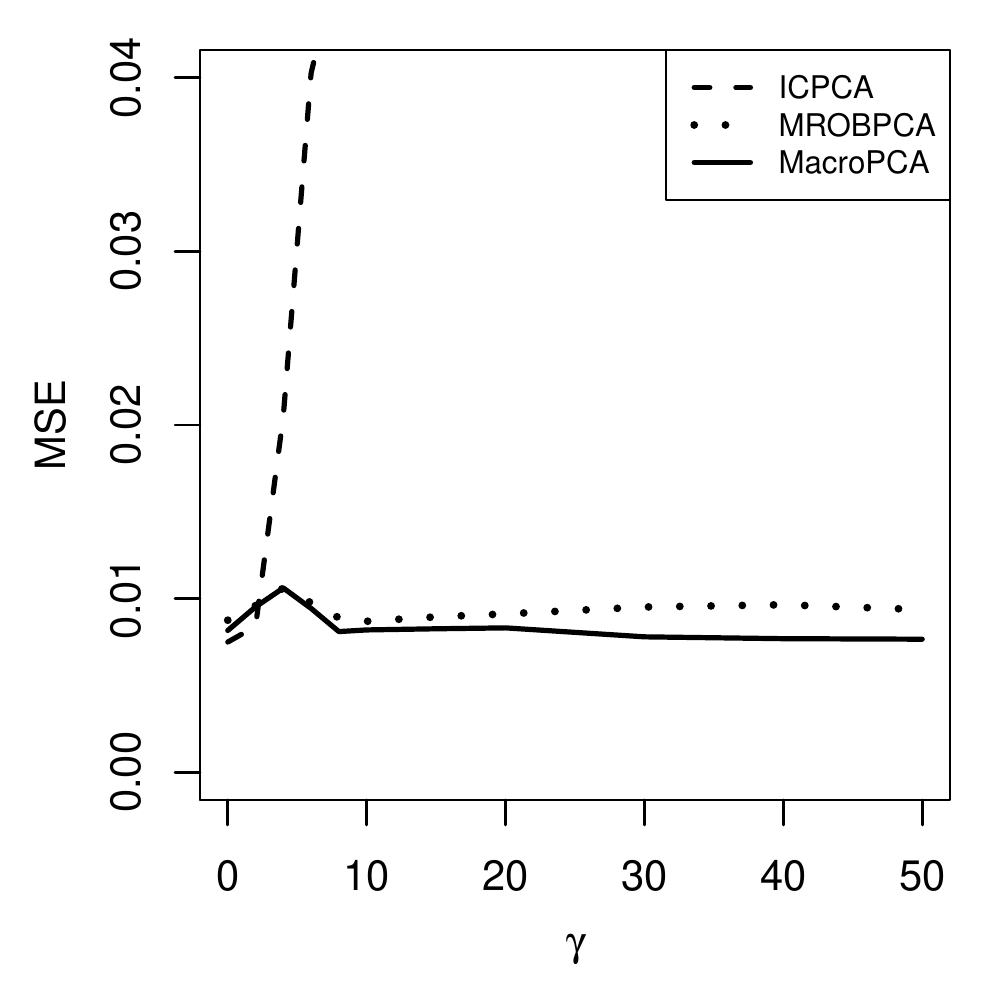} & 
		\includegraphics[width=0.45\textwidth]
	{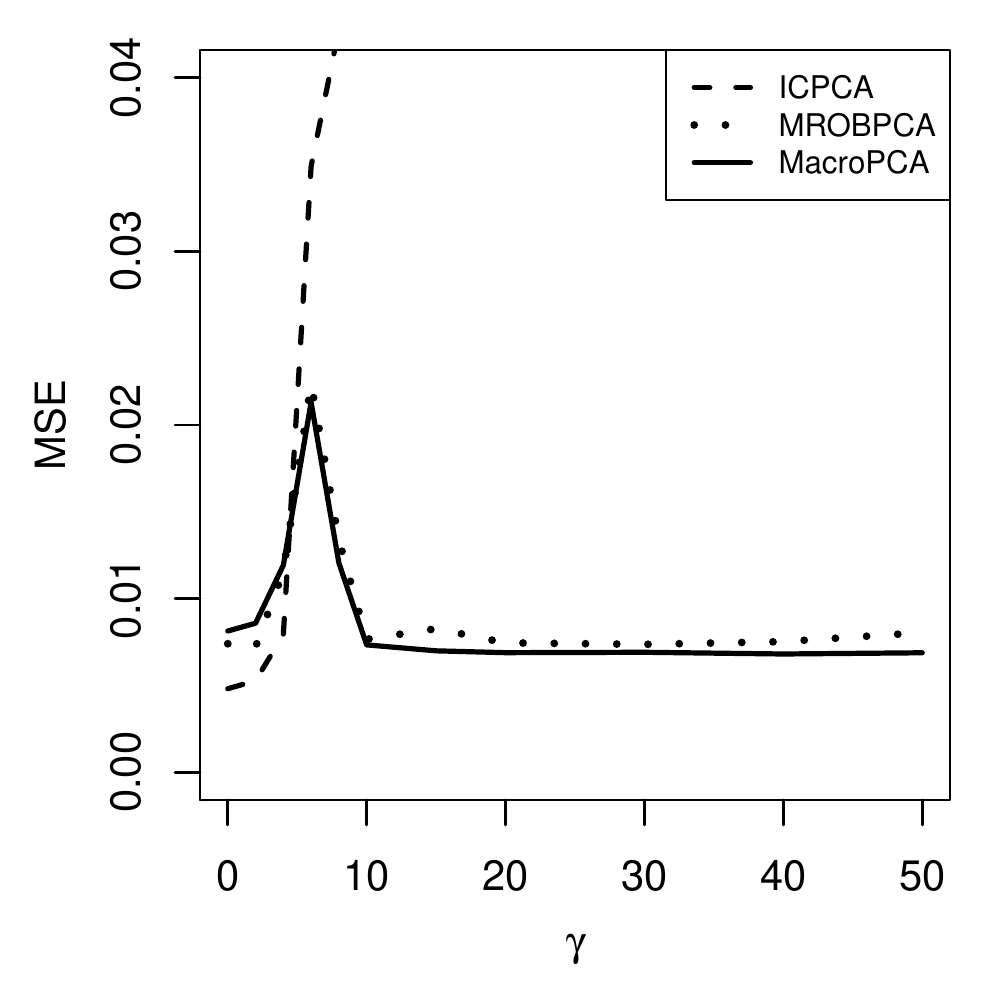}
\vspace{-0.2cm}
\end{tabular}
\caption{Average MSE for data with 20\% of MAR
    NA's and 20\% of rowwise outliers, as a 
		function of $\gamma$ which determines the 
		distance of the rowwise	outliers.} 
\label{fig:THCMMAR}
\end{figure}

\begin{figure}[!ht]
\centering
\vskip8mm
\begin{tabular}{cc}
\hskip 1mm A09, missing \& cellwise \& rowwise & 
\hskip 1mm ALYZ, missing \& cellwise \& rowwise \\
	\includegraphics[width=0.45\textwidth]
	{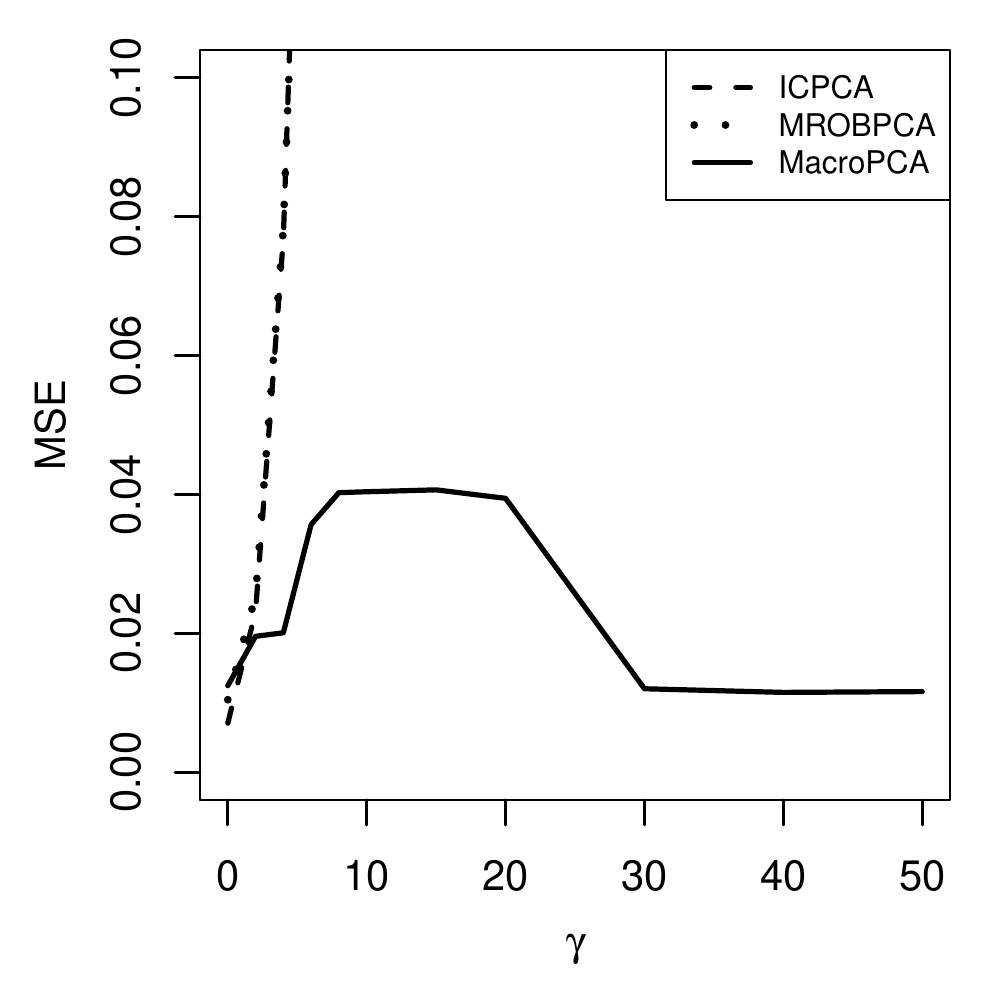} & 
	\includegraphics[width=0.45\textwidth]
	{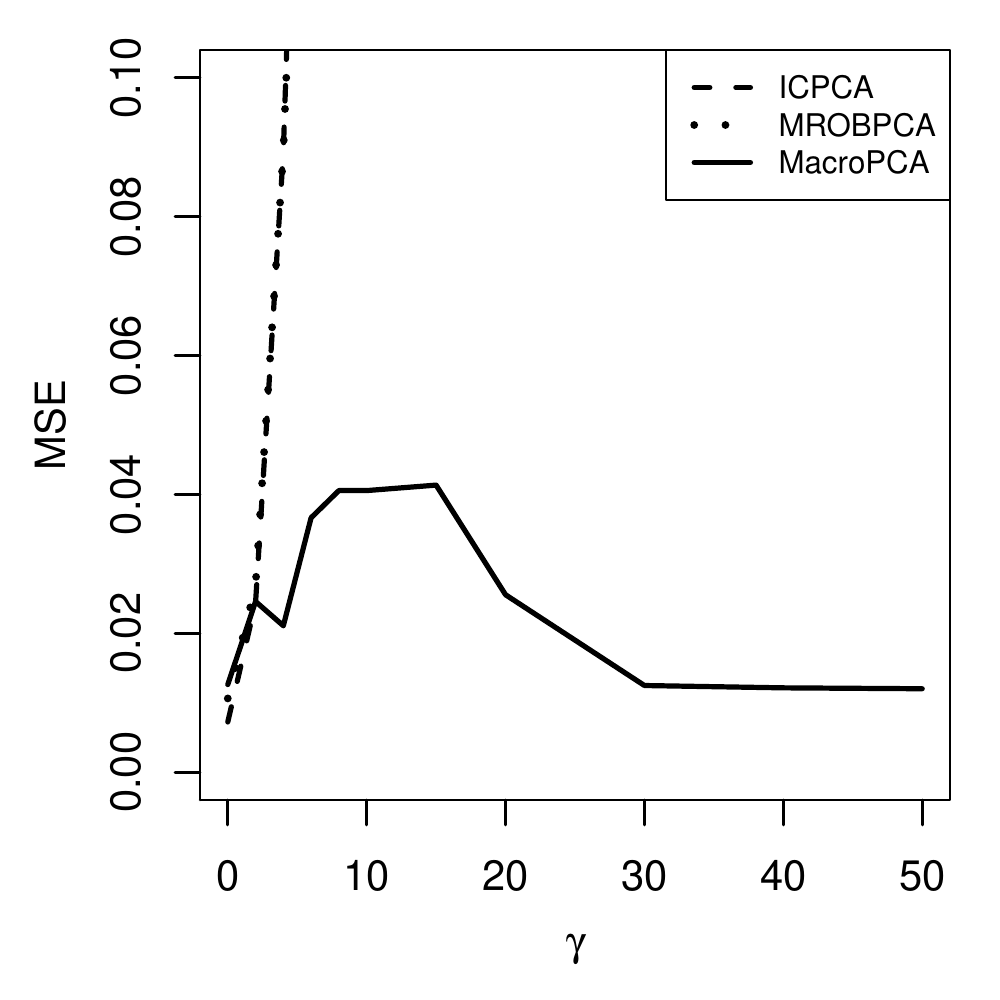}
\vspace{-0.2cm}
\end{tabular}
\caption{Average MSE for data with 20\% of MAR
    missing values, 10\% of cellwise outliers 
		and 10\% of rowwise outliers, 
		as a function of $\gamma$ 
		which determines the distance of both the 
		cellwise and the rowwise outliers.} 
\label{fig:BOTHMAR}
\end{figure}

Next, the other three settings in 
Section \ref{sec:Simulations} 
were repeated with MAR missing values.
This yielded Figures \ref{fig:ICMMAR},
\ref{fig:THCMMAR} and  \ref{fig:BOTHMAR},
which are extremely similar to the
corresponding Figures \ref{fig:ICM}, 
\ref{fig:THCM} and \ref{fig:BOTH}.
This confirms the suitability of MacroPCA 
in the MAR setting. 
		
\subsection{Simulation with data resembling DPOSS}
\label{A:DPOSS}

The DPOSS data analyzed in Subsection 
\ref{sec:DPOSS} has many NA's, in fact 50\%
of the cells are missing. 
The simulation study of Section 
\ref{sec:Simulations} did not include such
an extreme situation.
In order to check whether MacroPCA can handle
data with these characteristics, we redid the
simulation leading to Figure~\ref{fig:BOTH}
for $d=21$ variables and 50\% of missing values 
as well as 5\% of cellwise outliers and 5\% of 
rowwise outliers, all as in the DPOSS data.

\begin{figure}[!ht]
\centering
\vskip2mm
\begin{tabular}{cc}
\hskip 1mm A09, missing \& cellwise \& rowwise & 
\hskip 1mm ALYZ, missing \& cellwise \& rowwise \\
	\includegraphics[width=0.45\textwidth]
	{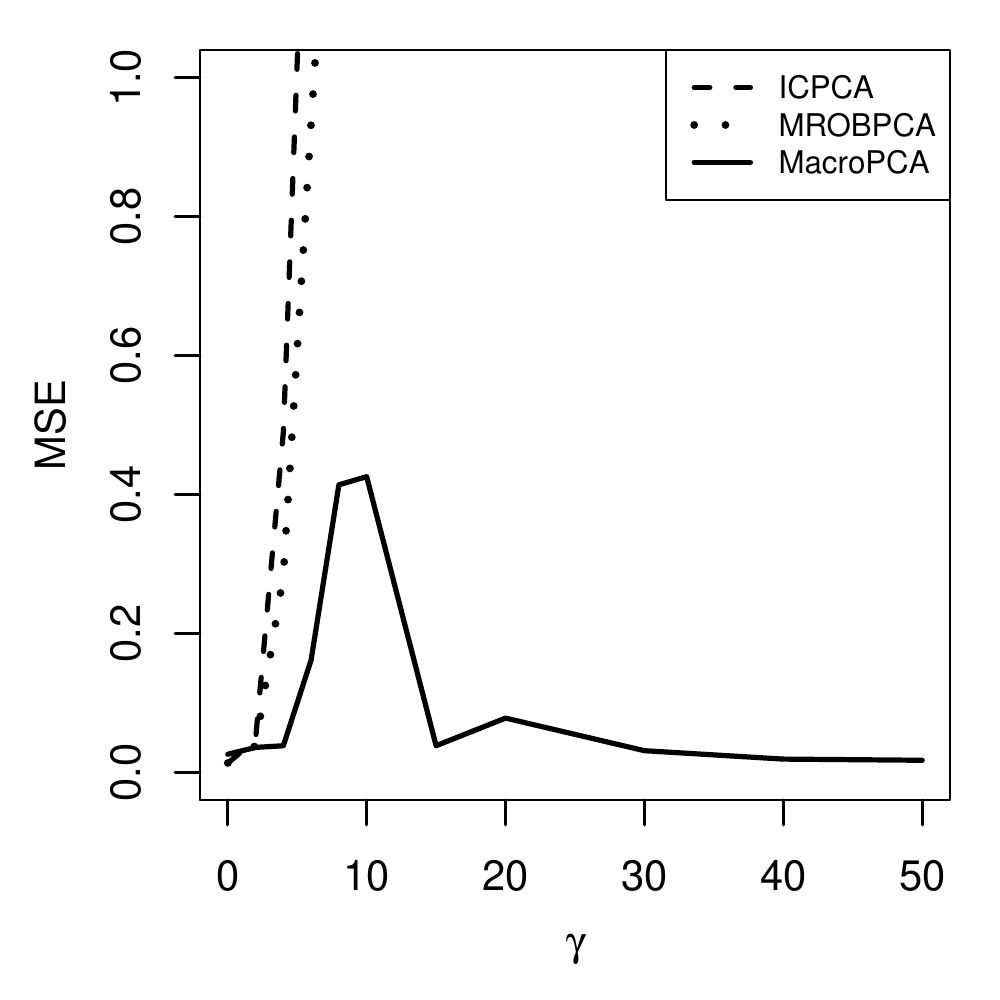} & 
	\includegraphics[width=0.45\textwidth]
	{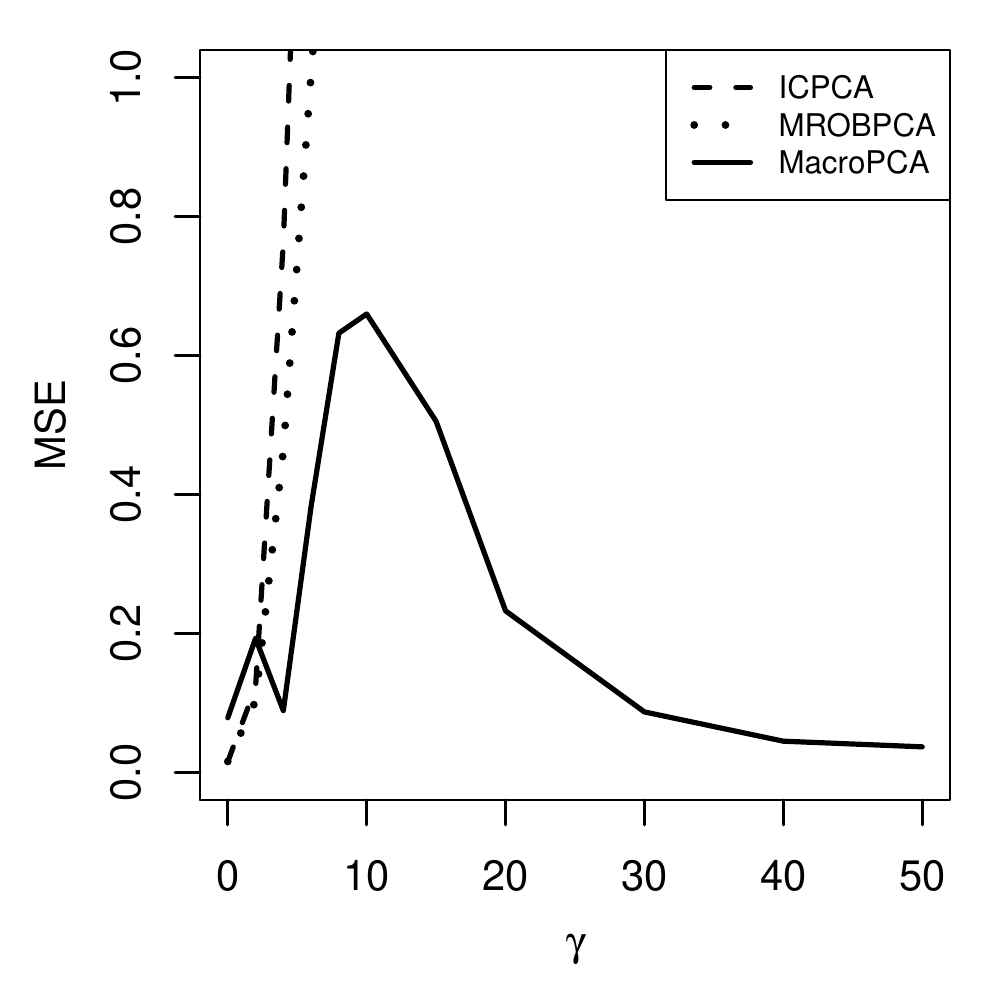}
\vspace{-0.2cm}
\end{tabular}
\caption{Average MSE for data with 21 variables,
    50\% of missing values, 5\% of cellwise outliers
		and 5\% of rowwise outliers, as a function of 
		$\gamma$ which determines the distance of 
		both the cellwise and the rowwise outliers.} 
\label{fig:missing50}
\end{figure}

Figure~\ref{fig:missing50} shows the results,
which indicate that ICPCA and MROBPCA broke down 
whereas MacroPCA still worked well.

\vspace{0.5cm}
\noindent {\bf Acknowledgments.}
The research of P. Rousseeuw has been supported 
by projects of Internal Funds KU Leuven.
W. Van den Bossche obtained financial support 
from the EU Horizon 2020 project SCISSOR: 
Security in trusted SCADA and 
smart-grids 2015-2018. 
The reviewers provided helpful suggestions
to improve the presentation.

\vspace{0.5cm}
\noindent {\bf Supplementary Material.} This is
a text with more details about the first part
of the MacroPCA algorithm and a list of notations.

\bibliographystyle{Chicago}



\clearpage
\pagenumbering{arabic}

\noindent{\Large{\bf Supplementary Material}}\\

\noindent{\bf 1. Description of the DDC algorithm}\\

Here we summarize the steps of the DDC algorithm, and 
refer to \citet{Rousseeuw:DDC} for more details. 

\begin{enumerate}
\item {\bf Standardization.}  
The location and scale estimates of each column $j$ 
of $\bX_{n,d}$ are calculated as
  $m_j = \textit{robLoc}_i(x_{ij})$ and 
	$s_j = \textit{robScale}_i(x_{ij} - m_j)$
where {\it robLoc} and {\it robScale} are 1-step 
M-estimators of location and scale. 
Then $\bX_{n,d}$ is standardized by column to 
$\bZ_{n,d}$ by $z_{ij} = (x_{ij}-m_j)/s_j \;$.

\item {\bf Univariate outlier detection.}
An initial cellwise outlier detection is performed
by flagging cells that are outlying in their column. 
To this end a new matrix $\bU_{n,d}$ is created in 
which univariate cellwise outliers are replaced by 
missing values, i.e. $\bU_{n,d}$ contains the entries
\begin{equation*}
  u_{ij} = \left\{
	\begin{array}{ll}
	z_{ij} & \,\,\, 
	\mbox{  if $|z_{ij}| \leqslant c_u$}\\
	\mbox{NA} & \,\,\, \mbox{  if $|z_{ij}| > c_u$}\\
	\end{array}
  \right.
\end{equation*}
where the cutoff value $c_u$ is set to 
$c_u = \sqrt{\chi^2_{1,p}}$ with probability $p=0.99$ 
by default. 

\item {\bf Bivariate relations.}
To reduce the propagation effect of cellwise outliers, 
only bivariate relations are considered.
For any two columns $j \neq l$ of $\bU_{n,d}$ we
calculate
   $\textit{cor}_{jl} = 
	  \textit{robCorr}_i(u_{ij},u_{il})$
where \textit{robCorr} is a robust correlation measure 
that discards missing values. Variables $j$ that 
satisfy $|\textit{cor}_{jl}| \geqslant 0.5$ for some
$l \neq j$ are called \textit{connected} while the 
others are called \textit{standalone} variables. 

The connected variables are sufficiently correlated 
to help predict each other. 
To this end we compute the robust slopes of the 
connected variables as
   $b_{jl} = \textit{robSlope}_i(u_{ij} | u_{il})$
where \textit{robSlope} robustly estimates the slope 
of a no-intercept regression line that predicts 
variable $j$ from variable $l$. 
Also the function \textit{robSlope} discards missing 
values.
 
\item {\bf Prediction.}
A crucial aspect of the DDC algorithm is its ability 
to robustly predict cell values. 
For the standalone variables all univariately outlying 
$z_{ij}$ are predicted by zero, since no further 
information is available.
(This means the unstandardized $x_{ij}$ are replaced
by the robust location estimate $m_j$\;.)
The non-outlying cell values are $z_{ij}$ are
predicted by themselves. 
The prediction of a connected variable is more involved. 
In words, for such a variable a set of `simple'
predictions is made, each using the slope of a variable 
to which it is connected. 
The final predictions are obtained by combining 
these simple predictions, using the correlations of 
the connected variables as weights. 

More formally, for each variable $j$ the set $C_j$ is 
considered which consists of all variables $l$ with
$|\textit{cor}_{jl}| \geqslant 0.5$, including $j$ 
itself. Next, for all $i=1,\ldots,n$ and $j=1,\ldots,d$ 
the predicted values are calculated as
\begin{equation}\label{eq:predz}\tag{S.1}
   \hat{z}_{ij} = 
	 \frac{\sum_{l}\; w_{jl}\,b_{jl}\,u_{il}} 
 	      {\sum_{l}\; w_{jl}} \; ,    
\end{equation}
where $w_{jl} = |cor_{jl}|$. 
The weighted average is taken over all $l \in C_j$ for 
which $u_{il}$ is not missing. 
Any missing value in $\bhZ_{n,d}$ is set to zero. 
Note that for standalone variables $\hat{z}_{ij}=u_{ij}$ 
since $C_j=\{j\}$.

\item {\bf Deshrinkage.}
Let us consider a column $j$.
The predictions $\hat{z}_{ij}$ typically have a smaller 
scale than that of the original column of $z_{ij}$. 
To compensate for this shrinkage, $\hat{z}_{ij}$ is 
replaced by
 $a_j \hat{z}_{ij}$ 	where 
 $a_j = \textit{robSlope}_{i'}(z_{i'j}|\hat{z}_{i'j}).$

\item {\bf Flagging cellwise outliers.}
The final predictions $\hat{z}_{ij}$ can be used to flag 
cellwise outliers. 
Any cell whose value differs too much from its prediction 
value is flagged. 
More precisely, the standardized cell residuals
\begin{equation} \label{eq:stdresDDC} \tag{S.2}
   r_{ij}^{(0)} = \frac{z_{ij} - \hat{z}_{ij}}
	  {\textit{robScale}_{i'}(z_{i'j} - \hat{z}_{i'j})} 
\end{equation}
are computed for all non-missing $z_{ij}$ and all cells 
with $|r_{ij}^{(0)}| > c_u$ are flagged.
Their indices (positions) are stored in a set 
$I_{c,DDC}$\;. 

\item {\bf Flagging rowwise outliers.}
The DDC method can also flag some outlying rows $i$ based 
on the standardized cell residuals $r_{ij}^{(0)}$\;. 
For multivariate Gaussian data without outliers 
$r_{ij}^{(0)} \approx N(0,1)$  so the cdf of 
$(r_{ij}^{(0)})^2$ is approximately the cdf 
$F$ of $\chi^2_1$. This motivates the criterion
\begin{equation} \label{eq:rowcrit} \tag{S.3}
   T_i = \ave_{j=1}^d F\left(\left(r_{ij}^{(0)}
	       \right)^2\right)	\;\;.
\end{equation} 
Next, the $T_i$ are standardized robustly and the rows 
$i$ for which the squared standardized $T_i$ exceeds 
the cutoff $c_u^2$ are flagged and stored 
in a set $I_{r,DDC}$\;. 

\item {\bf Imputation.}	
The NA-imputed matrix $\bcZ_{n,d}$ is assembled by 
replacing all missing values of the original $\bZ_{n,d}$ 
matrix with their predictions. The more fully imputed 
matrix $\btZ_{n,d}$ also replaces all cellwise outliers, 
i.e.
\begin{equation*}
  \tz_{i,j} = \left\{
	\begin{array}{ll}
		\hat{z}_{ij} & \,\,\, 
		\mbox{  if $|r_{ij}^{(0)}| > c_u$ or $z_{ij}$ is NA}\\
		z_{ij} & \,\,\, \mbox{  otherwise.} \\
	\end{array}
  \right.
\end{equation*}
These imputed matrices are turned into imputed 
matrices $\bcX_{n,d}$ and $\btX_{n,d}$ by undoing the 
standardization.
\end{enumerate}

Table S.1 lists all matrices, vectors and index sets 
used in this first stage of MacroPCA. 
The DDC algorithm has been implemented as the function
{\it DetectDeviatingCells} in the R package 
{\it cellWise} \citep{Raymaekers:cellWise} available 
in CRAN.

\newpage
\begin{table}[!ht]
\centering
Table S.1: Overview of notations used in the DDC 
 stage of MacroPCA\\
\vspace{0.3cm}
\begin{tabular}{ll}
\hline
$\bX_{n,d}$ & original data matrix, 
              with rows $\bx_i$ \\
$\bZ_{n,d}$ & standardized data matrix\\
$\bU_{n,d}$ & standardized data matrix with 
              univariate outliers set to NA \\
$\bhZ_{n,d}$ & matrix with predicted values for 
               $\bZ_{n,d}$ \\
$\bR_{n,d}^{(0)}$ & standardized residual matrix \\
$I_{c, DDC}$ & set of cells flagged as cellwise 
               outliers \\
$T_i$        & measure of rowwise outlyingness based 
               on $\bR_{n,d}^{(0)}$ \\
$I_{r,DDC}$  & set of rows flagged as possible rowwise 
               outliers, based on $T_i$ \\ 
$\bcZ_{n,d}$ & NA-imputed standardized data: all missing 
               cells replaced by\\
						 & predicted values \\
$\btZ_{n,d}$ & imputed standardized data matrix: all 
               cells in $I_{c,DDC}$ and all\\
						 & missing values imputed\\
$\bcX_{n,d}$ & NA-imputed data matrix,
               with rows $\bcx_i$\\ 
$\btX_{n,d}$ & imputed data matrix (all missing values 
               and all flagged cells are\\
						 &	imputed), with rows $\btx_i$\\ 
\hline
\end{tabular}
\end{table}

\vspace{1.5cm}
\noindent{\bf 2. Notations used in MacroPCA}\\

\indent The notations used in the second part of 
the MacroPCA algorithm are listed in Table S.2.

\begin{table}
\centering
Table S.2: Overview of notations used in the second 
stage of MacroPCA\\
\vspace{0.3cm}
\begin{tabular}{ll}
\hline
$\bfX^{(0)}_{n,d}$ & initial cell-imputed data matrix,
                     with rows $\bfx^{(0)}_i$\\
$H_0$ & index set of the $h$ rows with smallest 
        outlyingness \\
$k$   & dimension of the PCA subspace \\
$\bm_{d}^{(s)},\, \bL_{k,k}^{(s)}\,, \bP_{d,k}^{(s)}$ & 
              PCA estimates from iteration step $s$,
				      based on the imputed\\
				      & rows in $H_0$ \\
$\bcX_{n,d}^{(s)}$ & NA-imputed matrix (all missing 
                     cells imputed),
										 with rows $\bcx_i^{(s)}$\\
$\bfX_{n,d}^{(s)}$ & cell-imputed matrix (imputes 
                     outlying cells in rows of $H_0$\\ 
									 & and all missing values),
									   with rows $\bfx_i^{(s)}$\\
$\bhX_{n,d}^{(s)}$ & prediction for $\bfX^{(s)}$ based 
                     on $\bm_{d}^{(s-1)}$ and 
										 $\bP_{d,k}^{(s-1)}$,
										 with rows $\bhx_i^{(s)}$\\
$\bfT_{n,k}^{(s)}$ & scores of $\bfX^{(s)}$ with respect 
                     to $\bm_{d}^{(s)}$ and 
										 $\bP_{d,k}^{(s)}$ \\			
$\fod_i$ & orthogonal distance of $\bfx_i^{(s)}$ with 
         respect to the current PCA\\
			   & subspace \\
$H^*$ & index set of the $n^*$ rows with small
         enough $\fod$\\
$\bfX_{n,d}$ & cell-imputed data matrix at the end of 
               the algorithm, with\\
						 & rows $\bfx_i$\\
$\bm_d^*\,, \bP_{d,k}^*$ & center and PCA loadings of 
                             $\bfX_{n^*,d}$ \\
$\bfT_{n^*,k}$ & scores of the $n^*$ rows in 
                 $\bfX_{n^*,d}$ with respect to 
								 $\bm_{d}^*$ and $\bP_{d,k}^*$ \\
$\bm_k^{\mcd}\,, \bP_{k,k}^{\mcd}$ & 
                 robust center and loadings 
								 of $\bfT_{n^*,k}$ \\
$\bm_d\,, \bP_{d,k}$ & center and loadings of the final 
                         PCA fit\\
$\bcT_{n,k}$  & scores of all $n$ rows of
                $\bcX_{n,d}$ with respect to the final
								PCA fit \\ 
$\bchX_{n,d}$ & predicted values of all $n$ rows 
                of $\bcX_{n,d}\,$, 
								with rows $\bchx_i$ \\
$\btT_{n,k}$ & scores of all $n$ rows of $\btX_{n,d}$ with 
               respect to the final PCA fit \\ 
$\cod_i$ & orthogonal distance of $\bcx_i$ with respect 
         to the final PCA\\
			 & subspace \\
$\bR_{n,d}$ & final standardized residual matrix of $\bX$.\\
\hline
\end{tabular}
\end{table}

\end{document}